\documentclass[10pt,preprint]{aastex}

\slugcomment{}

\shorttitle{Gamma-ray emission from stellar-mass black holes}
\shortauthors{Hirotani et al.}

\begin{document}

%% LaTeX will automatically break titles if they run longer than
%% one line. However, you may use \\ to force a line break if
%% you desire.

\title{High-energy and Very-high-energy emission from 
       stellar-mass black holes moving in gaseous clouds}

%% Use \author, \affil, and the \and command to format
%% author and affiliation information.
%% Note that \email has replaced the old \authoremail command
%% from AASTeX v4.0. You can use \email to mark an email address
%% anywhere in the paper, not just in the front matter.
%% As in the title, use \\ to force line breaks.

% \author{Kouichi Hirotani\altaffilmark{1}, Hung-Yi Pu\altaffilmark{1}}
% \author{}
 \author{Kouichi Hirotani${}^1$, 
         Hung-Yi Pu${}^2$, 
         Sabrina Outmani${}^3$, 
         Hsinhao Huang${}^4$, 
         Dawoon Kim${}^5$, 
         Yoogeun Song${}^{6,7}$
         Satoki Matsushita${}^1$, and
         Albert K. H Kong${}^8$,
         }
 \affil{${}^1$
       Academia Sinica, Institute of Astronomy and Astrophysics (ASIAA),
       PO Box 23-141, Taipei, Taiwan 10617, R.O.C.;
       hirotani@asiaa.sinica.edu.tw}
 \affil{${}^2$
       Perimeter Institute for Theoretical Physics, 
       31 Caroline Street North, Waterloo, ON, N2L 2Y5, Canada}
 \affil{${}^3$
       School of Physics \& Astronomy, Queen Mary University of London, 
       Mile End Road, London, E1 4NS, U.K.}
 \affil{${}^4$
       Department of Physics, National Taiwan University, 
       No. 1 Sec. 4, Roosevelt Road, Taipei 10617, Taiwan}
 \affil{${}^5$
       Department of Physics, Dankook University, Cheonan 31116, 
       Republic of Korea}
 \affil{${}^6$
       Korea Astronomy and Space Science Institute, 
       Daejeon 305-348, Republic of Korea}
 \affil{${}^7$
       University of Science and Technology, 
       Daejeon 305-350, Republic of Korea}
 \affil{${}^8$
       Institute of Astronomy,
       National Tsing Hua University,
       No. 101, Section 2, Kuang-Fu Road, Hsinchu, Taiwan 30013, R.O.C.}

% \email{hirotani@tiara.sinica.edu.tw}

% \author{C. D. Biemesderfer\altaffilmark{4,5}}
% \affil{National Optical Astronomy Observatories, Tucson, AZ 85719}
% \email{aastex-help@aas.org}
% 
% \and
% 
% \author{R. J. Hanisch\altaffilmark{5}}
% \affil{Space Telescope Science Institute, Baltimore, MD 21218}
% 
%% Notice that each of these authors has alternate affiliations, which
%% are identified by the \altaffilmark after each name.  Specify alternate
%% affiliation information with \altaffiltext, with one command per each
%% affiliation.
% 
% \altaffiltext{1}{Visiting Astronomer, Cerro Tololo Inter-American Observatory.
% CTIO is operated by AURA, Inc.\ under contract to the National Science
% Foundation.}
% \altaffiltext{2}{Society of Fellows, Harvard University.}
% \altaffiltext{3}{present address: Center for Astrophysics,
%     60 Garden Street, Cambridge, MA 02138}
% \altaffiltext{4}{Visiting Programmer, Space Telescope Science Institute}
% \altaffiltext{5}{Patron, Alonso's Bar and Grill}

%% Mark off your abstract in the ``abstract'' environment. In the manuscript
%% style, abstract will output a Received/Accepted line after the
%% title and affiliation information. No date will appear since the author
%% does not have this information. The dates will be filled in by the
%% editorial office after submission.

\begin{abstract}
We investigate the electron-positron pair cascade taking place
in the magnetosphere of a rapidly rotating black hole.
Because of the spacetime frame dragging,
the Goldreich-Julian charge density changes sign
in the vicinity of the event horizon,
which leads to an occurrence of a magnetic-field aligned electric field,
in the same way as the pulsar outer-magnetospheric accelerator.
In this lepton accelerator, electrons and positrons 
are accelerated in the opposite directions,
to emit copious gamma-rays via 
the curvature and inverse-Compton processes.
We examine a stationary pair cascade,
and show that a stellar-mass black hole moving in a gaseous cloud
can emit a detectable very-high-energy flux,
provided that the black hole is extremely rotating
and that the distance is less than about 1~kpc.
We argue that the gamma-ray image will have a point-like morphology,
and demonstrate that their gamma-ray spectra have 
a broad peak around 0.01--1~GeV and a sharp peak around 0.1~TeV,
that the accelerators become most luminous 
when the mass accretion rate becomes about 0.01\% of the Eddington rate,
and that the predicted gamma-ray flux little changes
in a wide range of magnetospheric currents.
An implication of the stability of such a stationary gap is discussed.
\end{abstract}

%% Keywords should appear after the \end{abstract} command. The uncommented
%% example has been keyed in ApJ style. See the instructions to authors
%% for the journal to which you are submitting your paper to determine
%% what keyword punctuation is appropriate.


%% From the front matter, we move on to the body of the paper.
%% In the first two sections, notice the use of the natbib \citep
%% and \citet commands to identify citations.  The citations are
%% tied to the reference list via symbolic KEYs. The KEY corresponds
%% to the KEY in the \bibitem in the reference list below. We have
%% chosen the first three characters of the first author's name plus
%% the last two numeral of the year of publication as our KEY for
%% each reference.

%% Authors who wish to have the most important objects in their paper
%% linked in the electronic edition to a data center may do so by tagging
%% their objects with \objectname{} or \object{}.  Each macro takes the
%% object name as its required argument. The optional, square-bracket 
%% argument should be used in cases where the data center identification
%% differs from what is to be printed in the paper.  The text appearing 
%% in curly braces is what will appear in print in the published paper. 
%% If the object name is recognized by the data centers, it will be linked
%% in the electronic edition to the object data available at the data centers  
%%
%% Note that for sources with brackets in their names, e.g. [WEG2004] 14h-090,
%% the brackets must be escaped with backslashes when used in the first
%% square-bracket argument, for instance, \object[\[WEG2004\] 14h-090]{90}).
%%  Otherwise, LaTeX will issue an error. 

\section{Introduction}
\label{sec:intro}
% 
% A focal problem today in the dynamics of globular clusters is core collapse.  
% It has been predicted by theory for decades \citep{hen61,lyn68,spi85}, but
% observation has been less alert to the phenomenon. For many years the
% central brightness peak in M15 \citep{kin75,new78} seemed a unique anomaly.  
% Then \citet{aur82} suggested a central peak in \object{NGC 6397}, and a 
% limited photographic survey of ours \citep[Paper I]{djo84} found three more 
% cases, \objectname{NGC 6624}, \objectname[M 15]{NGC 7078}, and 
% \object[Cl 1938-341]{Terzan 8}), whose sharp center had often been 
% remarked on \citep{can78}.  
% 
By the Imaging Atmospheric Cherenkov Telescopes (IACTs),
76 very-high-energy (VHE) gamma-ray sources have been found
on the Galactic Plane\footnote{TeV Catalog (http:www.tevcat.uchicado.edu)}.
So far, 19 of them have been identified as pulsar wind nebulae,
10 of them as supernova remnants adjacent to molecular clouds,
whereas 36 of them are still remained unidentified.
To consider the nature of such unidentified VHE sources
in TeV energies,
a hadronic cosmic-ray cascade model has been proposed
\citep{ginz64,blandford87}.
In this model, charged particles such as protons
are accelerated in the blast waves of a supernova remnant (SNR)
% and propagate through the interstellar medium
and enter a dense molecular cloud.
Then proton-proton collisions take place,
leading to subsequent $\pi^0$ decays.
The resultant $\gamma$-rays will show
a single power-law spectrum between GeV and 100~TeV,
reflecting the energy distribution of the parent cosmic rays.
Its VHE emission morphology will become extended
and the centroid of the VHE image will be located close to the
peak of the gas density.
Thus, if a VHE source positionally coincides with a dense 
molecular cloud with an extended morphology,
and if the spectrum shows a single power-law between GeV and 100~TeV,
it strongly suggest that the emission is due to the $\pi^0$ decay
resulting from $p$-$p$ collisions.

On the other hand,
there is an alternative scenario for a VHE emission
from the magnetosphere of a rotating black hole (BH).
This BH lepton accelerator model, or the BH-gap model,
was first proposed by \citet{bes92}.
Then \citet{hiro98,nero07,rieger08,levi11,globus14,brod15,hiro16a,levi17}
extended this pioneering work and quantified
the BH-gap models.
In the present paper, 
we proposed that a BH gap is activated 
when a rapidly rotating BH enters a molecular cloud
or a gaseous cloud,
and that its maximum possible $\gamma$-ray fluxes
can be observable with the near-future IACTs such as the CTA,
provided that the stellar-mass BH is extremely rotating
and its distance is within 1~kpc.
The morphology of such a gap emission is predicted to be point-like,
and its spectrum will show two peaks around 0.1~GeV and 0.1~TeV.

On these grounds, 
to discriminate the physical origin of the VHE emissions,
it is essential to examine 
if the source is extended or point-like,
if the VHE peak coincides with the molecular density peak or not, and
if the $\gamma$-ray spectrum is power-law or bimodal.
We therefore briefly describe the observations of individual TeV sources
in the next section.
Then we describe the interactions between a BH
and a gaseous cloud in \S~\ref{sec:BH_cloud},
a stationary BH gap model in \S~\ref{sec:gap},
and the results in \S~\ref{sec:BH_gap_results}.
In the final section, we highlight the difference of the present model
from alternative $\gamma$-ray emission models from dense molecular clouds,
and discuss the electrodynamical stability of stationary BH gap solutions.

\section{Very-high-energy gamma-ray observations of the galactic plane}
\label{sec:VHE_obs}
In this section, we describe the VHE observations of 
individual sources along the galactic plane,
focusing on those associated with molecular clouds or unidentified.

It was pointed out that the VHE emission from HESS~J1457-593
positionally coincides with a giant molecular cloud (GMC) complex,
which overlaps the southern rim of SNR~G318.2+01
with the typical $\mbox{H}_2$ number density of $40 \mbox{ cm}^{-3}$
 \citep{hofver10}.
A two-dimensional Gaussian fit gives 
the source size of $\sigma_1=0.31^\circ$ and $\sigma_2=0.17^\circ$
along the major and minor axes, respectively.
However, the source has a non-Gaussian morphology,
which is likely further decomposed into 
two compact or point-like components in the north-south direction.
Here, we define that a TeV source is {\it compact} if
its angular size, $\sigma \equiv \sqrt{\sigma_1 \sigma_2}$,
is smaller than the angular resolution ($\sim 0.1^\circ$) of 
the present IACTs like the H.~E.~S.~S., 
and define that a TeV source is {\it point-like} if
$\sigma <0.5^\circ$. 
Therefore, if we observe the source with
the new IACTs, Cherenkov Telescope Array (CTA),
we may be able to decompose HESS~J1457-593 into northern and southern
compact and/or point-like components, 
both of which coincide with the peaks of 
${}^{12}\mbox{CO}(J:1\rightarrow 0)$ line emission
in the southern part of SNR G318.2+01.
Interestingly, another dense molecular cloud exits about $0.5^\circ$
west of HESS~J1457-593 and overlaps the southern rim of 
SNR G318.2+01; 
however, this dense molecular cloud does not show any detectable
TeV emissions.
It may be due to the propagation effect of the cosmic rays
in the SNR shell;
however, it may be due to a coincidental passage of 
one or two BHs in the GMC that positionally 
coincides with HESS~J1457-593.

Subsequently, \citet{aharon08a,aharon08b,aharon08c}
reported the positional coincidence of six TeV sources,
HESS~J1714-385, HESS~J1745-303, 
HESS~J1801-233, HESS~J1800-240A, B, and C
with dense molecular clouds.
HESS~J1714-385 has a compact morphology with size
$\sigma=0.07^\circ$ and positionally associated with an extended
dense molecular clouds whose density is 
$  1.5 \times 10^2 \mbox{ cm}^{-3} < n_{{\rm H}_2} 
 < 6.6 \times 10^2 \mbox{ cm}^{-3}$.
However, the peak of the TeV emission resides in the valley 
between the two $\mbox{H}_2$ density peaks.
This positional deviation from the molecular density peak 
may be due to the propagation effect of the CRs emitted from
SNR~37A, or may be due to a passage of a BH
in the molecular cloud.
HESS~J1745-303 has an extended morphology
and show the VHE emission above 20~TeV;
thus, we consider that the VHE photons are emitted 
via a hadronic interaction between CRs and the molecular clouds
for this source.
HESS~J1801-233, HESS~J1800-2400A and B are extended
and roughly overlaps the density peaks of the molecular cloud
whose averaged molecular density is 
$n_{{\rm H}_2} \sim 10^3 \mbox{ cm}^{-3}$.
Thus, VHE photons may be emitted by the interaction between CRs and
molecular clouds in these three TeV sources.
The remaining one TeV source, HESS~J1800-2400C,
has a point-like morphology with $\sigma=0.02^\circ$ 
and appear to be deviated from the peak of the molecular density.
In short, among these six TeV sources that are positionally 
associated with dense molecular clouds, 
two sources, HESS~J1714-385 and HESS~J1800-2400C,
have compact and point-like morphology, respectively.
It is noteworthy that the centroids of their TeV emission 
deviate from the nearby peaks of molecular hydrogen column density.
Therefore, if their emission morphology is found to be point-like with CTA,
and if the VHE spectrum cuts off around 1~TeV 
(see \S~\ref{sec:results_SED}),
the BH-gap scenario may account for these two VHE sources. 

Following these pioneering works mentioned just above,
% \cite{hofver10,aharon08a,aharon08b,aharon08c}, 
\citet{deWilt17} carried out a systematic comparison between 
TeV sources and dense molecular gas along the galactic plane.
They used published HESS data up to 2015 March
and picked up 49 TeV sources with 11--15~mm radio observations
of molecular emission lines.
They found that 38 of the 49 sources are positionally
associated with dense gas counterparts;
specifically speaking, $\mbox{NH}_3\mbox{(1,1)}$ line emissions were
detected from or adjacent to the 38 TeV sources.
Moreover, out of unidentified 18 TeV sources,
12 of them are positionally associated with dense molecular clouds.
Among these 12 TeV sources, 9 sources were fit with Gaussian model,
5 of which are found to have compact morphology.
Specifically, 
HESS~J1634-472, 
% HESS~J1702-420, HESS~J1708-410,
HESS~J1804-216, and HESS~J1834-087
have the sizes of $\sigma=0.11^\circ$, 
% $0.08^\circ$, $0.054^\circ$, 
$0.20^\circ$, and $0.09^\circ$, respectively \citep{aharon06};
thus, one of the three sources is compact.
% where J1702-420 and J1708-410 are excluded to avoid
% an overlap with \citep{aharon08d}.
Also,
HESS~J1472-608, HESS~J1626-490, HESS~J1702-420, HESS~J1708-410, 
HESS~J1841-055, % HESS~J1857+026, and HESS~J1858+020
have 
$\sigma=0.056^\circ$, $0.083^\circ$, $0.212^\circ$, $0.069^\circ$, and
$0.320^\circ$, % $0.093^\circ$, and $0.040^\circ$, 
respectively \citep{aharon08d};
thus, three of the five sources are compact.
Subsequently, 
HESS~J1641-463 is also found to be compact with $\sigma=0.085^\circ$
\citep{abramo14}. 
For the 38 TeV sources positionally associated with dense molecular
clouds, the molecular hydrogen density is typically in the range
$10^3 \mbox{ cm}^{-3} < n_{{\rm H}_2} < 10^5 \mbox{ cm}^{-3}$.
% (positional coincidence/deviation with/from MC?)
Among the 5 compact TeV sources that are positionally associated
with dense molecular clouds,
the centroid of HESS~J1626-490 coincides with
the molecular density peak;
thus, this source may be due to the interaction between CRs and 
molecular clouds.
However, there is a possibility that BH gaps emit the observed
TeV photons for the remaining 4 compact sources,
in addition to the point-like source HESS~J1800-2400C.
In what follows, we thus investigate if a stellar-mass BH
can emit detectable TeV photons when they enter a gaseous cloud.

\section{Black holes moving in a gaseous cloud}
\label{sec:BH_cloud}
Before proceeding to the BH gap model,
we must consider the mass accretion process
when a BH moves in a dense gaseous cloud.
Thus, in the subsequent three subsections,
we briefly describe the giant molecular clouds (GMCs),
formation of BHs, and the accretion process in a gaseous cloud.

\subsection{Giant molecular clouds}
\label{sec:GMC}
Molecular clouds are generally gravitationally bound
and occasionally contain several sites of star formation.
Particularly, massive stars formed in a GMC can ionize 
the surrounding interstellar medium with their strong UV radiation.
A combined action of such ionization,
stellar winds, and supernova explosions,
blow off the gases in a GMC,
leaving an OB association adjacent to dense molecular clouds.
In this section, we focus on the physical parameters
in such dense molecular clouds,
which may be traversed by a BH formed in a
neighboring OB association. 

The physical parameters (e.g., temperature and density)
of a molecular cloud can be examined
by observing the strength, width, and profile of radio emission lines 
of probe molecules.
A typical GMC has gas kinematic temperature
between $30$~K and $50$~K.
Using the dense-gas tracers, ${\rm H}_2{\rm CO}$ and CS,
we can infer the hydrogen molecule number density, 
$n_{{\rm H}_2}$ in the core of a GMC. 
A typical GMC core has the density
$10^4 \mbox{ cm}^{-3} < n_{{\rm H}_2} < 10^6 \mbox{ cm}^{-3}$,
and mass between $10M_\odot$ and $10^3 M_\odot$.
Individual molecular clouds have lower densities,
$10^2 \mbox{ cm}^{-3} < n_{{\rm H}_2} < 10^5 \mbox{ cm}^{-3}$
and masses between $10^3 M_\odot$ and $10^6 M_\odot$.
We use these values of $n_{{\rm H}_2}$
to estimate the mass accretion rate onto a BH
in \S~\ref{sec:Bondi}.

\subsection
{Black hole formation in GMCs}
\label{sec:OBassoc}
To consider the passage of a stellar-mass BH in a 
dense molecular cloud, let us briefly comment on 
the massive star formation in a GMC.
In a GMC,
massive stars are formed in OB associations.
A typical OB association contains
$10^{1-2}$ high-mass stars of type O and B
and $10^{2-3}$ stars of lower masses.
The stellar line-of-sight velocity dispersion 
in an OB association is
typically around $9 \mbox{ km s}^{-1}$
and usually less than $20  \mbox{ km s}^{-1}$
\citep{sitnik03}.
The strong winds and the supernovae 
resulting from these associations,
blow off the interstellar medium; 
thus, OB associations are found adjacent to molecular clouds
\citep{blitz80}.
Depending on the mass and metalicity, 
such high-mass, OB stars evolve into neutrons stars or BH
after core collapse events.
For example,
if the progenitor has mass $M$ in $25M_\odot < M < 40 M_\odot$
with low or solar metalicity,
it will evolve into a BH through a supernova explosion
after the fall back of material onto an initial neutron star.
In this case, the BH will acquire a certain kick velocity
with respect to the star-forming region 
in the similar way as neutron stars.
On the other hand,
if it has $M>40M_\odot$ with low metalicity,
it will evolve into a BH directly 
without a supernova explosion.
These massive BHs will have smaller
relative velocities with respect to the star-forming region
compared to the lighter BHs formed through 
core-collapse supernovae.
Nevertheless, we may expect that such BHs,
whichever formed with or without supernovae, 
move into nearby dense molecular clouds.
Thus, in the next subsection,
we estimate the plasma accretion rate 
when such a stellar-mass BH enters a dense gaseous cloud.

\subsection
{Bondi accretion rate in the molecular cloud core}
\label{sec:Bondi}
To estimate the accretion rate onto a BH,
we examine the Bondi accretion rate 
when a BH moves in a gaseous cloud.
To this end, we begin with comparing the sound speed
in a typical molecular cloud and the relative velocity
between the cloud and the BH.

The sound speed of a cloud with kinetic temperature $T$ (K)
can be estimated by
\begin{equation}
  C_{\rm s} \sim \sqrt{\frac{k_{\rm B} T}{m_{\rm p}}}
           = 90 \, T^{1/2} \mbox{ m s}^{-1},
  \label{eq:S1}
\end{equation}
where 
$k_{\rm B}$ refers to the Boltzmann constant, and
$m_p$ the proton mass.
For a typical GMC, we have $T < 50$~K,
and for a typical dark cloud, we have $T < 20$~K.
Thus, we obtain $C_{\rm s} < 630 \mbox{ m s}^{-1}$
as the upper limit of the sound speed in a molecular cloud.

A typical velocity $V$ of a BH
relative to the cloud may be estimated by the kick velocity
in a supernova explosion.
For a neutron star, it is typically a few hundred kilometers per second.
For a BH, a greater fraction of mass is expected to be turned into
a compact object; so, we may expect the typical velocity is around
$10^2 \mbox{ km s}^{-1}$.
Turbulent velocities measured from molecular line width are
usually less than $10 \mbox{ km s}^{-1}$, and
relative velocities among smaller scale clouds are also 
within this small range.
Thus, we can neglect such random or bulk motions 
and adopt the supernova kick velocity, 
$V \sim 10^2 \mbox{ km s}^{-1}$,
as the typical velocity of a BH with respect to the molecular clouds.
If a heavier BH ($M>40M_\odot$) 
is formed without a supernova explosion, 
the relative velocity will be less than this value.

On these grounds, we can safely put $V \gg C_{\rm s}$.
In this case, the flow becomes supersonic with respect to the BH
and a shock wave is formed behind the hole.
Accordingly, the gas particles within 
the Bondi radius (i.e., within the impact parameter)
$r_{\rm B} \sim GM/V^2$
from the BH, are captured,
falling onto the BH with the Bondi accretion rate,
\citep{bondi44}
\begin{equation}
  \dot{M}_{\rm B} = 4\pi \lambda (GM)^2 V^{-3} \rho,
  \label{eq:S2}
\end{equation}
where $\rho$ denotes the mass density of the gas,
and $\lambda$ is a constant of order unity.
We have $\lambda=1.12$ and $0.25$ for an isothermal 
and adiabatic gas, respectively.
Assuming molecular hydrogen gas,
and normalizing with the Eddington accretion rate,
we obtain the following dimensionless Bondi accretion rate,
\begin{equation}
  \dot{m}_{\rm B} 
  = 5.39 \times 10^{-9} \lambda n_{{\rm H}_2} M_1
    \left(\frac{\eta}{0.1}\right)^{-1}
    \left(\frac{V}{10^2 \mbox{ km s}^{-1}}\right)^{-3},
  \label{eq:S3}
\end{equation}
where $n_{{\rm H}_2}$ is measured in $\mbox{cm}^{-3}$ unit,
$\eta \sim 0.1$ denotes the radiation efficiency of the accretion flow,
and $M_1 \equiv M/(10 M_\odot)$.

Since the accreting gases have little angular momentum as a whole 
with respect to the BH, they form an accretion disk only 
within a radius that is much less than $r_{\rm B}$. 
Thus, we neglect the mass loss as a disk wind 
between $r_{\rm B}$ and the inner-most region, 
and evaluate the accretion rate near the BH,
$\dot{m}$, with $\dot{m}_{\rm B}$.
As will be shown in \S~\ref{sec:results_SED},
the gap of a stellar-mass BH becomes most luminous when
$6 \times 10^{-5} < \dot{m} < 2 \times 10^{-4}$.
For $\dot{m}$ to reside in this range,
a dense, isothermal molecular cloud core should have a density
$n_{{\rm H}_2} > 10^4 \mbox{ cm}^{-3}$,
if $V=100 \mbox{km s}^{-1}$.
If $V=50~\mbox{ km s}^{-1}$, however,
a lower density, $n_{{\rm H}_2} > 1.2 \times 10^{3} \mbox{ cm}^{-3}$,
is enough to activate the BH gap.

\section{Magnetospheric lepton accelerator model}
\label{sec:gap}
In this section, we formulate the BH gap model
and examine the resultant gamma-ray emission
when a stellar-mass BH moves in a dense molecular cloud.
We quickly review the pulsar outer gap model
in section~\ref{sec:OG}, 
and apply it to BH magnetospheres
in section~\ref{sec:BH_gap}, 
% in S~3.2. % \ref{sec:BH_gap}, 
focusing on the improvements from previous works by the authors.

\subsection
{Pulsar outer-magnetospheric lepton accelerator model}
\label{sec:OG}
The Large Area Telescope (LAT) aboard the
 {\it Fermi} space gamma-ray observatory
has detected pulsed signals
in high-energy (0.1~GeV-10~GeV) gamma-rays
from more than 200 rotation-powered pulsars
\footnote{Public List of LAT-Detected Gamma-Ray Pulsars
  (https://confluence.slac.stanford.edu/display/GLAMCOG/Public+List+of+LAT-Deteced+Gamma-Ray+Pulsars)}. 
Among them, 
20 pulsars exhibit pulsed signals above 10~GeV, 
including 10 pulsars up to 25~GeV
and other 2 pulsars above 50~GeV.
Moreover, more than 99\% of the LAT-detected 
young and millisecond pulsars
exhibit phase-averaged spectra that are consistent
with a pure-exponential or a sub-exponential cut off 
above the cut-off energies at a few GeV.
What is more, 30\% of these young pulsars
show sub-exponential cut off,
a slower decay than the pure-exponential functional form.
These facts preclude the possibility
of emissions from the inner magnetosphere 
as in the polar-cap scenario 
\citep{harding78,daugherty82,dermer94,timokhin13,timokhin15},
which predicts super-exponential cut off
due to magnetic attenuation.
That is, we can conclude that 
the pulsed emissions are mainly emitted from the
outer magnetosphere, 
which is close to or outside the light cylinder.

One of the main scenarios of such outer-magnetospheric
emissions is the outer-gap model 
\citep{cheng86a,cheng86b,chiang92,romani96,cheng00,hiro06b}.
In the present paper, 
we apply this successful scenario to BH magnetospheres.
Although the electrodynamics is mostly common between
the pulsar outer-gap model and the present BH-gap model,
there is a striking difference between them.
In a pulsar magnetosphere, 
an outer gap arises because of the convex geometry of the
dipolar-like magnetic field in the outer magnetosphere.
However, in a BH magnetosphere,
a gap arises because of the frame-dragging 
in the vicinity of the event horizon.
We describe this BH-gap model below.

% \paragraph*{Black hole gap model} 
%
\subsection
{Black-hole inner-magnetospheric lepton accelerator model}
\label{sec:BH_gap}
\subsubsection{Background spacetime geometry}
\label{sec:geometry}
In a rotating BH magnetosphere,
electron-positron accelerator is formed in the direct vicinity
of the event horizon.
Thus, we start with describing the background spacetime
in a fully general-relativistic way.
We adopt the geometrized unit, putting $c=G=1$, where
$c$ and $G$ denote the speed of light and the gravitational constant,
respectively.
Around a rotating BH, 
the spacetime geometry is described by the Kerr metric
\citep{kerr63}.
% becomes axisymmetric with respect to the
% rotation axis (with $\theta=0$) and is 
In the Boyer-Lindquist coordinates, it becomes 
\citep{boyer67} % ,misner73}
\begin{equation}
 ds^2= g_{tt} dt^2
      +2g_{t\varphi} dt d\varphi
      +g_{\varphi\varphi} d\varphi^2
      +g_{rr} dr^2
      +g_{\theta\theta} d\theta^2,
  \label{eq:metric}
%  \eqno{(S1)},
\end{equation}
where 
\begin{equation}
   g_{tt} 
   \equiv 
   -\frac{\Delta-a^2\sin^2\theta}{\Sigma},
   \qquad
   g_{t\varphi}
   \equiv 
   -\frac{2Mar \sin^2\theta}{\Sigma}, 
  \label{eq:metric_2}
%  \eqno{(S2)}
\end{equation}
\begin{equation}
   g_{\varphi\varphi}
     \equiv 
     \frac{A \sin^2\theta}{\Sigma} , 
     \qquad
   g_{rr}
     \equiv 
     \frac{\Sigma}{\Delta} , 
     \qquad
   g_{\theta\theta}
     \equiv 
     \Sigma ;
  \label{eq:metric_3}
%  \eqno{(S1)}
\end{equation}
$\Delta \equiv r^2-2Mr+a^2$,
$\Sigma\equiv r^2 +a^2\cos^2\theta$,
$A \equiv (r^2+a^2)^2-\Delta a^2\sin^2\theta$.
At the horizon, we obtain $\Delta=0$, 
which gives the horizon radius, 
$r_{\rm H} \equiv M+\sqrt{M^2-a^2}$,
where $M$ corresponds to the gravitational radius,
$r_{\rm g} \equiv GM c^{-2}=M$.
The spin parameter $a$ becomes $a=M$ for a maximally rotating BH,
and becomes $a=0$ for a non-rotating BH. 
The spacetime dragging frequency is given by
$\omega(r,\theta)= -g_{t\varphi}/g_{\varphi\varphi}$,
which decreases outwards as $\omega \propto r^{-3}$
at $r \gg r_{\rm g}=M$.

\subsubsection{Poisson equation for the non-corotational potential}
\label{sec:poisson}
We assume that the non-corotational potential $\Phi$
depends on $t$ and $\varphi$ only through
the form $\varphi-\Omega_{\rm F} t$, and put
\begin{equation}
  F_{\mu t}+\Omega_{\rm F} F_{\mu \varphi}
  = -\partial_\mu \Phi(r,\theta,\varphi-\Omega_{\rm F} t) ,
  \label{eq:def_Phi}
\end{equation}
where $\Omega_{\rm F}$ denotes the magnetic-field-line
rotational angular frequency.
We refer to such a solution as a \lq stationary' solution
in the present paper.

The Gauss's law gives the Poisson equation
that describes $\Phi$ in a three dimensional magnetosphere
\citep{hiro06b},
\begin{equation}
  -\frac{1}{\sqrt{-g}}
   \partial_\mu 
      \left( \frac{\sqrt{-g}}{\rho_{\rm w}^2}
             g^{\mu\nu} g_{\varphi\varphi}
             \partial_\nu \Phi
      \right)
  = 4\pi(\rho-\rho_{{\rm GJ}}),
  \label{eq:pois}
\end{equation}
where the general-relativistic Goldreich-Julian (GJ) charge density
is defined as
\citep{hiro06b}
\begin{equation}
  \rho_{\rm GJ} \equiv 
      \frac{1}{4\pi\sqrt{-g}}
      \partial_\mu \left[ \frac{\sqrt{-g}}{\rho_{\rm w}^2}
                         g^{\mu\nu} g_{\varphi\varphi}
                         (\Omega_{\rm F}-\omega) F_{\varphi\nu}
                 \right].
  \label{eq:def_GJ}
\end{equation}
Far away from the horizon, $r \gg M$, 
equation~(\ref{eq:def_GJ}) reduces to the
ordinary, special-relativistic expression
of the GJ charge density~\citep{GJ69,mestel71},
\begin{equation}
  \rho_{\rm GJ} 
  \equiv -\frac{\mbox{\boldmath$\Omega$}\cdot\mbox{\boldmath$B$}}
               {2\pi c}
         +\frac{(\mbox{\boldmath$\Omega$}\times\mbox{\boldmath$r$})\cdot
                (\nabla\times\mbox{\boldmath$B$})}
               {4\pi c}.
  \label{eq:def_rhoGJ_1}
\end{equation}
Therefore, the corrections due to magnetospheric currents,
which are expressed by the second term of eq.~(\ref{eq:def_rhoGJ_1}),
are included in equation~(\ref{eq:def_GJ}).

If the real charge density $\rho$ deviates from the
rotationally induced GJ charge density,
$\rho_{\rm GJ}$, in some region,
equation~(\ref{eq:pois}) shows that
$\Phi$ changes as a function of position.
Thus, an acceleration electric field, 
$E_\parallel= -\partial \Phi / \partial s$,
arises along the magnetic field line,
where $s$ denotes the distance along the magnetic field line.
A gap is defined as the spatial region in which $E_\parallel$
is non-vanishing.
At the null charge surface,
$\rho_{{\rm GJ}}$ changes sign by definition.
Thus, a vacuum gap, 
in which $\vert\rho\vert \ll \vert \rho_{{\rm GJ}} \vert$,
appears around the null-charge surface,
because $\partial E_\parallel / \partial s$
should have opposite signs at the inner and outer boundaries
\citep{chiang92,romani96,cheng00}. % ,HS99}.
As an extension of the vacuum gap, 
a non-vacuum gap,
in which $\vert\rho\vert$ becomes a good fraction of 
$\vert \rho_{{\rm GJ}} \vert$,
also appears around the null-charge surface
(\S~2.3.2 of HP~16),
unless the injected current across either the inner or the outer
boundary becomes a substantial fraction of the GJ value.
If the injected current becomes non-negligible compared to the created
current in the gap, the gap centroid position shifts from the 
null surface; however, 
the essential gap electrodynamics does not change.

In previous series of our papers
\citep{hiro16a,hiro16b,song17},
we have assumed $\Delta \ll M^2$ in Equation~(\ref{eq:pois}), 
expanding the left-hand side in the series of $\Delta/M^2$
and pick up only the leading orders.
However, in the present paper, 
we discard this approximation, 
and consider all the terms that arise
at $\Delta \sim M^2$ or $\Delta \gg M^2$.

It should be noted that
$\rho_{\rm GJ}$ vanishes, and hence the null surface appears
near the place where $\Omega_{\rm F}$ coincides with 
the space-time dragging angular frequency, $\omega$ \citep{bes92}.
The deviation of the null surface
from this $\omega(r,\theta)=\Omega_{\rm F}$ surface is,
indeed, small, as figure~1 of \citet{hiro98} indicates.
Since $\omega$ can match $\Omega_{\rm F}$ only near the horizon,
the null surface, and hence the gap generally appears 
within one or two gravitational radii above the horizon,
irrespective of the BH mass. 

% Include more basic eqs.

\subsubsection{Particle Boltzmann equations}
\label{sec:Boltzmann}
We outline the Boltzmann equations of $e^\pm$'s,
following \citet{hiro17}.
Imposing a stationary condition, 
$\partial/\partial t
  + \Omega_{\rm F} \partial/\partial \phi = 0,
  \label{eq:stationary}
$
we obtain the following Boltzmann equations, 
\begin{equation}
  c\cos\chi  \frac{\partial n_\pm}{\partial s}
  +\dot{p}   \frac{\partial n_\pm}{\partial p}
  = \alpha(S_{{\rm IC},\pm} +S_{{\rm p},\pm}),
 \label{eq:BASIC_2}
\end{equation}
along each radial magnetic field line on the poloidal plane,
where the upper and lower signs correspond to the
positrons (with charge $q=+e$) and electrons ($q=-e$), respectively, and
$p \equiv \vert\mbox{\boldmath$p$}\vert=m_{\rm e}c\sqrt{\gamma^2-1}$.
The left-hand side is in $dt$ basis, where $t$ denotes the
proper time of a distant static observer.
Thus, the lapse $\alpha$ is multiplied in the right-hand side,
because both $S_{\rm IC}$ and $S_{\rm p}$ are evaluated in 
the zero-angular-momentum observer (ZAMO).
Dimensionless lepton distribution functions 
per magnetic flux tube are defined by
\begin{equation}
  n_\pm \equiv 
    \frac{2\pi ce}{\Omega_{\rm F} B} N_\pm(r,\theta,\gamma),
  \label{eq:def_n}
\end{equation}
where $N_+$ and $N_-$ designate the 
distribution functions of positrons and electrons, respectively;
$\gamma$ refers to these lepton's Lorentz factor,
and $B \equiv \vert \mbox{\boldmath$B$} \vert$. 
% Dimensionless GJ charge density per magnetic flux tube
% is defined by
% \begin{equation}
%   n_{\rm GJ} \equiv \frac{2\pi c}{\Omega_{\rm F} B} \rhoGJ .
%   \label{eq:def_nGJ}
% \end{equation}
It is convenient to include the curvature emission
as a friction term in the left-hand side;
in this case, we obtain
\begin{equation}
  \dot{p} \equiv q E_\parallel \cos\chi -\frac{P_{\rm SC}}{c},
  \label{eq:char1}
\end{equation}
where the pitch angle is assumed to be
$\chi=0$ for outwardly moving positrons, and
$\chi=\pi$ for inwardly moving electrons.
The curvature radiation force 
is given by 
\citep[e.g.,][]{harding81}, % \citet{cheng96,zhang97}.
$P_{\rm SC}/c=2e^2 \gamma^4/(3 R_{\rm c}{}^2)$.

The IC collision terms are expressed as
\begin{eqnarray}
  S_{\rm IC} 
  &\equiv&
  -\int_{\epsilon_\gamma < \gamma}
     d \epsilon_\gamma
     \eta_{\rm IC}^\gamma(\epsilon_\gamma,\gamma,\mu_\pm) n_\pm
  \nonumber\\
  &+&
   \int_{\gamma_i > \gamma}
     d \gamma_i
     \eta_{\rm IC}^{\rm e}(\gamma_i,\gamma,\mu_\pm) n_\pm ,
  \label{eq:sec_ICa}
\end{eqnarray}
where the IC redistribution function is defined by
\begin{equation}
  \eta_{\rm IC}^\gamma % (\epsilon_\gamma,\gamma,\mu_\pm)
  \equiv
    (1-\beta\mu_\pm)
    \int_{E_{\rm min}}^{E_{\rm max}}
      dE_{\rm s} 
        \frac{dF_{\rm s}}{dE_{\rm s}}
        \frac{d\epsilon_\gamma^\ast}{d\epsilon_\gamma}
        \int_{-1}^{1} d\Omega_\gamma^\ast
          \frac{d\sigma_{\rm KN}^\ast}
               {d\epsilon_\gamma^\ast d\Omega_\gamma^\ast},
  \label{eq:src_ICb}
\end{equation}    
$m_{\rm e}c^2 \epsilon_\gamma$ denotes the upscattered $\gamma$-ray
energy.
The asterisk denotes that the quantity evaluated in the
electron (or positron) rest frame and 
$d\sigma_{\rm KN}^\ast/d\epsilon_\gamma^\ast d\Omega_\gamma^\ast$
denotes the Klein-Nishina differential cross section.
Energy conservation gives 
\begin{equation}
  \eta_{\rm IC}^{\rm e}(\gamma_i,\gamma,\mu_\pm)
  = \eta_{\rm IC}^\gamma(\gamma_i-\gamma,\gamma_i,\mu_\pm),  
\end{equation}
where $\gamma_i$ denotes the Lorentz factor before collision
and $\mu_+$ (or $\mu_-$) does 
the cosine of the collision angle with the
soft photon for outwardly moving electrons
(or inwardly moving positrons). 
For more details, see \S~3.2.2 of \citet{hiro03}.
%{\bf
The soft photons are emitted by the hot electrons within 
a radiatively inefficient accretion flow (RIAF). 
%}  % end of \bf
The effect of this inhomogeneous and anisotropic soft photon field
is included in the differential soft photon flux,
$dF_{\rm s}/dE_{\rm s}$,
through the correction factor $f_{\rm riaf}$
(\S~3.4 of \citet{hiro17}).
That is, we put
$dF_{\rm s}/dE_{\rm s}=f_{\rm riaf} \cdot (dF_{\rm s}/dE_{\rm s})_0$. 

The photon-photon pair creation term becomes
\begin{equation}
  S_{\rm p} \equiv
    \int d\nu_\gamma
      \alpha_{\gamma\gamma}
      \frac{2\pi e}{\Omega_{\rm F} B}
      \int \frac{I_\omega}{\hbar\omega} d\Omega_\gamma,
  \label{eq:src_1b}
\end{equation}
where 
\begin{equation}
  \alpha_{\gamma\gamma}
  = (1-\mu_\pm)
    \int_{E_{\rm th}}^\infty
        dE_{\rm s} 
        \frac{dF_{\rm s}}{dE_{\rm s}}
        \frac{d\sigma_{\gamma\gamma}}{d\gamma},
%        \frac{\partial^2 \sigma_{\gamma\gamma}}{d\gamma d\chi},
  \label{eq:def_alf2}
\end{equation}
The $\gamma$-ray specific intensity
$I_\omega$ is integrated over the $\gamma$-ray propagation
solid angle $\Omega_\gamma$.
For details, see \S~3.2.2 of \citep{hiro03}.
Note that $dF_{\rm s}/dE_{\rm s}$ 
in both $\eta_{\rm IC}^\gamma$ and $\alpha_{\gamma\gamma}$
is evaluated in ZAMO 
(\S~3.4 of \citet{hiro17}).

It is noteworthy that the charge conservation ensures that
the dimensionless total current density (per magnetic flux tube),
$j_{\rm tot} \equiv \int (-n_+ -n_-)d\gamma$ 
conserves along the flowline.
If we denote the created current density as $J_{\rm cr}$,
the injected current density across the inner and outer boundaries
as $J_{\rm in}$ and $J_{\rm out}$, respectively,
and the typical GJ value as
$J_{\rm GJ} \equiv \Omega_{\rm F}B_{\rm H}/(2\pi)$,
we obtain
$j_{\rm tot}= j_{\rm cr} + j_{\rm in} + j_{\rm out}$,
where 
$j_{\rm cr}  \equiv J_{\rm cr} /J_{\rm GJ}$,
$j_{\rm in}  \equiv J_{\rm in} /J_{\rm GJ}$,
$j_{\rm out} \equiv J_{\rm out}/J_{\rm GJ}$.

\subsubsection{Radiative transfer equation}
\label{sec:RTE}
In the same manner as \citep{hiro16b}, % \citep{hiro16b},
we assume that all photons are emitted with vanishing angular momenta
and hence propagate on a constant-$\theta$ cone.
Under this assumption of radial propagation,
we obtain the radiative transfer equation \citep{hiro13}, 
\begin{equation}
  \frac{d I_\omega}{dl}= -\alpha_\omega I_\omega + j_\omega,
  \label{eq:RTE}
\end{equation}
where $dl=\sqrt{g_{rr}}dr$ 
refers to the distance interval along the ray in ZAMO,
$\alpha_\omega$ and $j_\omega$ 
the absorption and emission coefficients
evaluated in ZAMO, respectively.
We consider both photon-photon and magnetic absorption,
pure curvature and IC processes for primary lepton emissions, and
synchrotron and IC processes for the emissions 
by secondary and higher-generation pairs.
For more details of the computation of 
absorption and emission,
see \S\S~4.2 and 4.3 of HP16
and \S~5.1.5 of H16. % \citep{hiro16}.

\subsubsection{Boundary conditions}
\label{sec:BDCs}
The elliptic type second-order partial differential 
equation~(\ref{eq:pois}) is solved on the 2-D poloidal plane.
We assume a reflection symmetry, 
$\partial_\theta \Phi=0$, at $\theta=0$.
We assume that the polar funnel is bounded at
a fixed colatitude, $\theta=\theta_{\rm max}$
and impose that this lower-latitude boundary
is equi-potential and put $\Phi=0$ 
at $\theta=\theta_{\rm max}=60^\circ$.
Both the outer and inner boundaries are treated as free boundaries.
At both inner and outer boundaries, 
$E_\parallel=-\partial \Phi / \partial r$ vanishes.

Since the magnetospheric current is to be constrained by a global
condition including the distant dissipative region,
we should treat $j_{\rm cr}$, $j_{\rm in}$ and $j_{\rm out}$ 
as free parameters, when we focus on the local gap electrodynamics.
For simplicity, 
we assume that there is no electron injection across the inner boundary
and put $j_{\rm in}=0$ throughout this paper.
In what follows, we examine stationary gap solutions for 
several representative values of $j_{\rm cr}$ and $j_{\rm out}$.

The radiative-transfer equation~(\ref{eq:RTE}),
a first-order ordinary differential, 
contains no photon injection across neither
the outer nor the inner boundaries.

\subsubsection{Gap closure condition}
\label{sec:closure}
We impose the same gap closure condition described
in \S~4.2.5 of \citet{hiro17}.
Namely, we impose 
${\cal M}_{\rm in} {\cal M}_{\rm out} = 1$,
where ${\cal M}_{\rm in}$ and ${\cal M}_{\rm out}$
denote the multiplicity of primary positrons and electrons,
respectively.
For more details, see \citet{hiro17}.

%{\bf
\subsubsection{Advection dominated accretion flow}
\label{sec:ADAF}
At a low accretion rate as discussed in \S~\ref{sec:Bondi}, 
the equatorial accretion flow becomes optically thin for 
Bremsstrahlung absorption and radiatively inefficient 
because of the weak Coulomb interaction between the ions and electrons.
% Accordingly, the ions stores the heat within the flow itself and 
% accretes onto the BH without losing the thermal energy as radiation .
This radiatively inefficient flow can be described by an 
advection-dominated accretion flow (ADAF)
\citep{narayan94,narayan95},
and provides the target soft photons for the IC-scattering 
and the photon-absorption processes in the polar funnel.
Thus, to tabulate the redistribution functions for these two processes,
we compute the specific intensity of the ADAF-emitted photons.
For this purpose, we adopt the analytical self-similar ADAF spectrum 
presented in \citet{mahad97}. 
The spectrum includes the contribution of the synchrotron, IC, 
and Bremsstrahlung processes.
These three cooling mechanisms balance with the heating 
due to the viscosity and the energy transport form ions,
and determine the temperature of the electrons in an ADAF 
to be around $T_{\rm e}\sim10^9$~K.
In radio wavelength, the ADAF radiation field 
is dominated by the synchrotron component 
whose peak frequency, $\nu_{\rm c,syn}$, 
varies with the accretion rate as $\nu_{\rm c,syn}\propto \dot{m}^{1/2}$. 
In X-ray wavelength, the Bremsstrahlung component
dominates the ADAF flux at such a low $\dot{m}$.
In soft $\gamma$-ray wavelength,
this component cuts off around the energy $h\nu\approx k T_{\rm e}$.
These MeV photons (with energies slightly below $kT_{\rm e}$)
collide each other to materialize as seed electrons and positrons
that initiate a pair-production cascade within the gap.
If the accretion rate is as low as $\dot{m} < 10^{-2.5}$,
the seed pair density becomes less than the GJ value
\citep{levi11},
thereby leading to an occurrence of a vacuum gap in the funnel.
However, if the accretion rate exceeds this critical value
and becomes $\dot{m} > 10^{-2.5}$,
the seed pair density exceeds the GJ value;
as a result, the magnetosphere becomes no longer charge-starved
and the gap ceases to exist.
%} % end of \bf

\section{Stationary BH gap solutions}
\label{sec:BH_gap_results}
We apply the method in the foregoing section to a stellar-mass
BH with mass $M=10 M_\odot$ and spin parameter $a=0.99 M$. 
Except for the BH mass and the surrounding environment,
the difference from \citet{hiro17,song17} appears in two major points.
First, we pick up all the terms in the left-hand side of
Equation~(\ref{eq:pois}),
discarding the approximation $\Delta \ll M^2$.
Second, consider a current injection across the outer boundary
in \S~\ref{sec:results_injected}.

\subsection{Gamma-ray emission from the black hole 
            moving in a gaseous cloud}
\label{sec:results_Epara}
Let us begin with the examination of the 
$E_\parallel(r,\theta)$ distribution 
along the individual magnetic field lines
that are radial on the meridional plane.
As demonstrated in figure~3 of \citet{hiro18},
$E_\parallel$ peaks along the rotation axis,
because the magnetic fluxes concentrate towards the
rotation axis
as the BH spin approaches its maximum value
(i.e., as $a \rightarrow r_{\rm g}$)
\citep{komissa07,tchekhov10}.
Therefore, to consider the greatest gamma-ray flux,
we focus on the emission along the rotation axis,
$\theta=0^\circ$.
The acceleration electric field, $E_\parallel$,
decreases slowly outside the null surface
in the same way as pulsar outer gaps \citep{hiro99}.
This is because the two-dimensional screening effect of $E_\parallel$
works when the gap longitudinal (i.e., radial) width
becomes non-negligible compared to its trans-field 
(i.e., meridional) thickness. 
In addition, in the present work, we pick up all the terms
that contribute not only near the horizon 
(i.e., $\Delta \ll M^2$)
but also away from it
(i.e., $\Delta \sim M^2$ or $\Delta \gg M^2$).
As a result, 
the exerted $E_\parallel$ is reduced from the case
of $\Delta \ll M^2$,
which particularly reduces the curvature luminosity 
compared to our previous works \citep{lin17}.

\begin{figure}
  \centering
     \includegraphics[width=8cm]{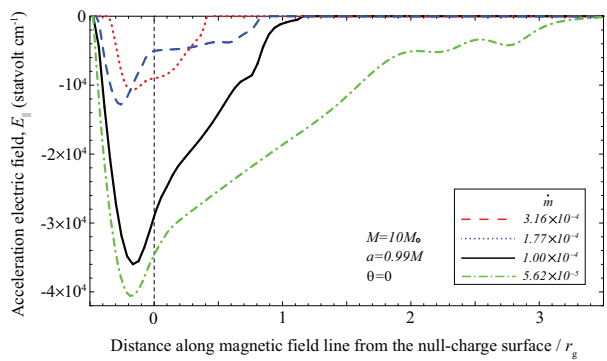}
  \caption{
Acceleration electric field, $E_\parallel(s)$,
along the rotation axis, $\theta=0$,
where $s$ denotes the distance (in Boyer-Lindquist radial coordinate)
from the null-charge surface along the poloidal magnetic field line.
The created current density is 70\% of the Goldreich-Julian value,
and the injected currents are set to be zero.
The red dashed, blue dotted, green dash-dotted, and black solid
curves corresponds to
$\dot{m}\equiv \dot{M}/\dot{M}_{\rm Edd}
 =3.16 \times 10^{-4}$, $1.77 \times 10^{-4}$, 
$1.00 \times 10^{-4}$, and $5.62 \times 10^{-5}$, respectively.
The vertical dashed line shows the position of the null-charge surface.
          }
\label{fig:Epara_mdot}
\end{figure}

In figure~\ref{fig:Epara_mdot}, 
we present $E_\parallel(s,\theta=0^\circ)$
solved at four dimensionless accretion rates,
$\dot{m}=10^{-3.50}$, $10^{-3.75}$, $10^{-4.00}$, $10^{-4.25}$,
where $s \equiv r-r_0(\theta)$ denotes the distance
from the null surface, $r=r_0(\theta)$,
and $\theta=0^\circ$ is adopted.
As pointed out in previous BH gap models,
the potential drop increases with decreasing $\dot{m}$.
However, if the accretion further decreases as 
$\dot{m} < 10^{-4.25}$,
there exists no stationary gap solutions.
Below this lower bound accretion rate,
the gap solution becomes inevitably non-stationary.
% The maximum potential drop for $\dot{m}=10^{-4.25}$
% (i.e., the green dash-dotted line) 
% results in the greatest gamma-ray flux (Fig.~3).
% Since the outwardly accelerated electrons attain the highest
% Lorentz factors for $\dot{m}=10^{-4.25}$ among these four cases,
% the resultant VHE spectrum becomes the hardest (Fig.~3).

% \begin{figure}
%   \centering
%      \includegraphics[width=8cm]{fig_1e1_a99_Ella.eps}
%   \caption{
% Distribution of the magnetic-field-aligned electric field, 
% $E_\parallel$, that is presented in figure~1,
% at four discrete colatitudes as labeled in the box,
% where $\theta=0^\circ$ corresponds to the rotation axis.
% The abscissa denotes the distance along the magnetic field
% from the null-charge surface,
% where the general relativistic Goldreich-Julian charge density
% vanishes due to the spacetime dragging around a rotating black hole.
% Black hole's mass ($M=10 M_\odot$),
% spin ($a=0.99r_{\rm g}$), 
% and the accretion rate ($\dot{m}=1.00 \times 10^{-4}$)
% are common with figure~1.
% The vertical dashed line shows the position of the null-charge surface
% along $\theta=0^\circ$;
% however, its position little depends on $\theta$
% because we assume $\Omega_{\rm F}=0.5 \omega_{\rm H}$.
%           }
% \label{fig:Epara}
% \end{figure}

\subsection
{Ultra-relativistic leptons}
\label{sec:results_distr}
\begin{figure}
  \centering
     \includegraphics[width=8cm]{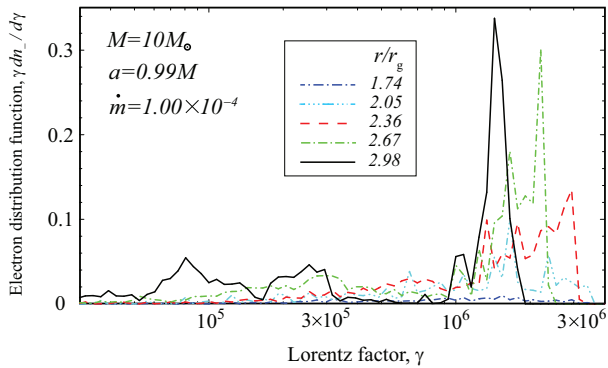}
  \caption{
Distribution function of the outwardly propagating electrons
at five discrete positions for the same BH parameters
as figures.~1. 
Electron Lorentz factors are saturated around $3 \times 10^6$ 
because of the curvature radiation drag. 
Electrons also distribute in the Lorentz factors 
below $(1.3 \sim 2) \times 10^6$
because of the inverse-Compton radiation drag
in the deep Klein-Nishina regime.
          }
\label{fig:distr}
\end{figure}

We next consider the electrons' distribution function.
Because of the the negative $E_\parallel$,
electrons and positrons are accelerated outward and inward, respectively.
As figure~\ref{fig:distr} shows, 
the electrons' Lorentz factors concentrate around 
$3 \times 10^6$ due to the curvature-radiation drag.
At the same time, electrons distribute at lower Lorentz factors
with a broad plateau typically between 
$6 \times 10^4$ and $2 \times 10^6$.
Electrons stay at such relatively lower Lorentz factors
because of the inverse-Compton drag.
Since the Klein-Nishina cross section increases with decreasing
Lorentz factors, such lower-energy electrons with
$6 \times 10^4 < \gamma < 2 \times 10^6$ efficiently contribute
to the VHE emission via the inverse-Compton scatterings.
% It results in a substantial VHE emission
% from a black-hole gap
% when the accretion rate approaches the lower bound,
% below which stationary gap solutions no longer exist.

\subsection{Gamma-ray spectra}
\label{sec:results_SED}
\begin{figure}
  \centering
     \includegraphics[width=8cm]{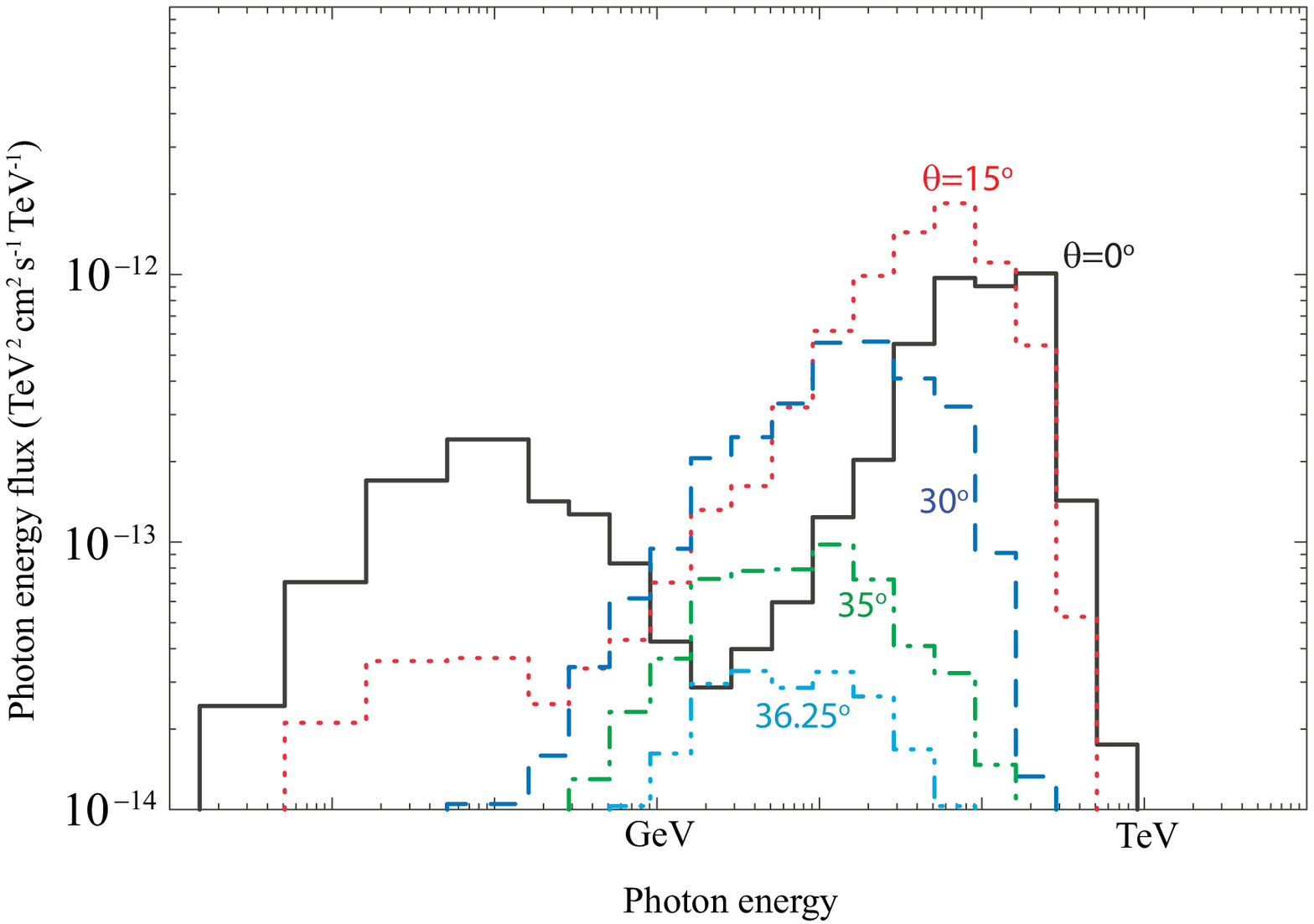}
  \caption{
Spectral energy distribution (SED) of the gap-emitted photons
along five discrete viewing angles with respect to the rotation axis, 
$\theta= 0^\circ$ (black solid line),
$\theta=15^\circ$ (red dotted),
$\theta=30^\circ$ (blue dashed),
$\theta=35^\circ$ (green dash-dotted), and
$\theta=36.25^\circ$ (cyan dash-dot-dot-dotted).
The BH mass and spin are chosen to be $M=10M_\odot$ and $a=0.99M$,
and the accretion rate is $\dot{m}=1.00 \times 10^{-4}$.
          }
\label{fig:SED_5th}
\end{figure}

Let us examine the gamma-ray spectra.
In figure~\ref{fig:SED_5th},
we present the Spectral energy distribution (SED)
of the gap emission
along five discrete viewing angles.
It follows that the gap luminosity maximizes
if we observe the BH almost face-on, $\theta < 15^\circ$,
and that the gap luminosity rapidly decreases
at $\theta < 30^\circ$ if the gap equatorial boundary
is located at $\theta=60^\circ$.
In what follows, to estimate the maximally possible VHE flux,
we consider the emission along the rotation axis, $\theta=0^\circ$.

\begin{figure}
  \centering
     \includegraphics[width=8cm]{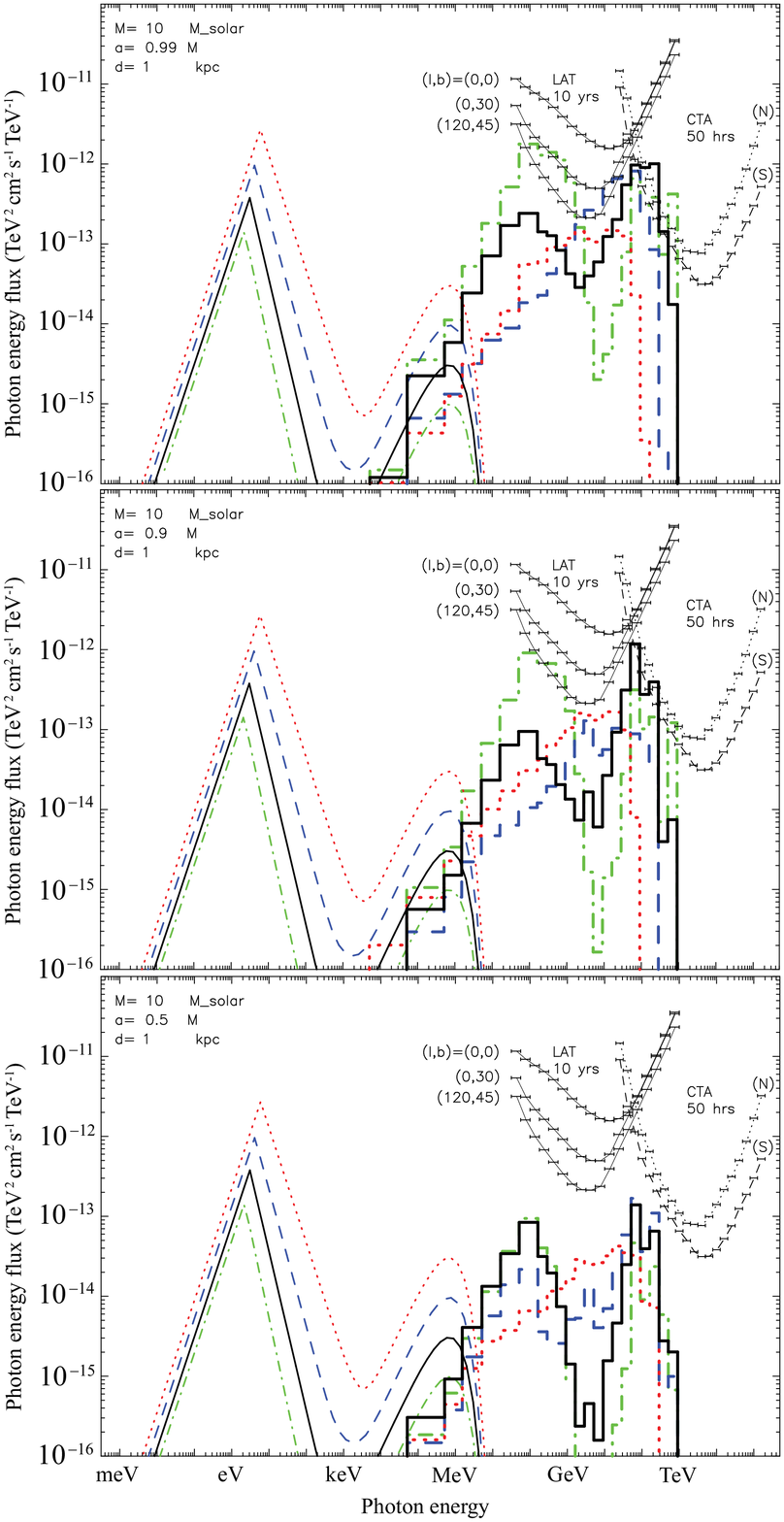}
  \caption{
SED of the gap-emitted photons
along the rotation axis, $\theta=0^\circ$, for $M=10M_\odot$.
The top panel shows the SED for $a=0.99M$, 
while the middle and bottom ones for $a=0.90M$ and $a=0.50M$,
respectively.
The created current density is 70\% of the Goldreich-Julian value,
and the injected currents are set to be zero.
The four thin curves in the left part of each panel
show the ADAF emission for 
$\dot{m}=3.16 \times 10^{-4}$ (red dotted), 
$1.77 \times 10^{-4}$ (blue dashed), 
$1.00 \times 10^{-4}$ (black solid), and 
$5.62 \times 10^{-5}$ (green dash-dotted).
The four thick curves in the right
show the BH gap emission for the corresponding $\dot{m}$.
          }
\label{fig:SED_spin}
\end{figure}

We also consider how the the SED depends on the BH spin.
In figure~\ref{fig:SED_spin},
we show the SEDs for $a=0.99M$, $0.90M$, and $0.50M$;
in each panel, SEDs for the four discrete accretion rates, 
$\dot{m}= 10^{-3.50}$, $10^{-3.75}$, $10^{-4.00}$, and $10^{-4.25}$,
are plotted.
It is clear that the gap luminosity increases with
decreasing $\dot{m}$.
It also follows that the gap emission could be detectable
with CTA if $a>0.90M$, 
provided that the distance is within 1~kpc and
we observe nearly face on.
However, if the BH is moderately rotating as $a=0.50M$,
it is very difficult to detect its emission,
unless it is located within 0.3~kpc.

\begin{figure}
  \centering
     \includegraphics[width=8cm]{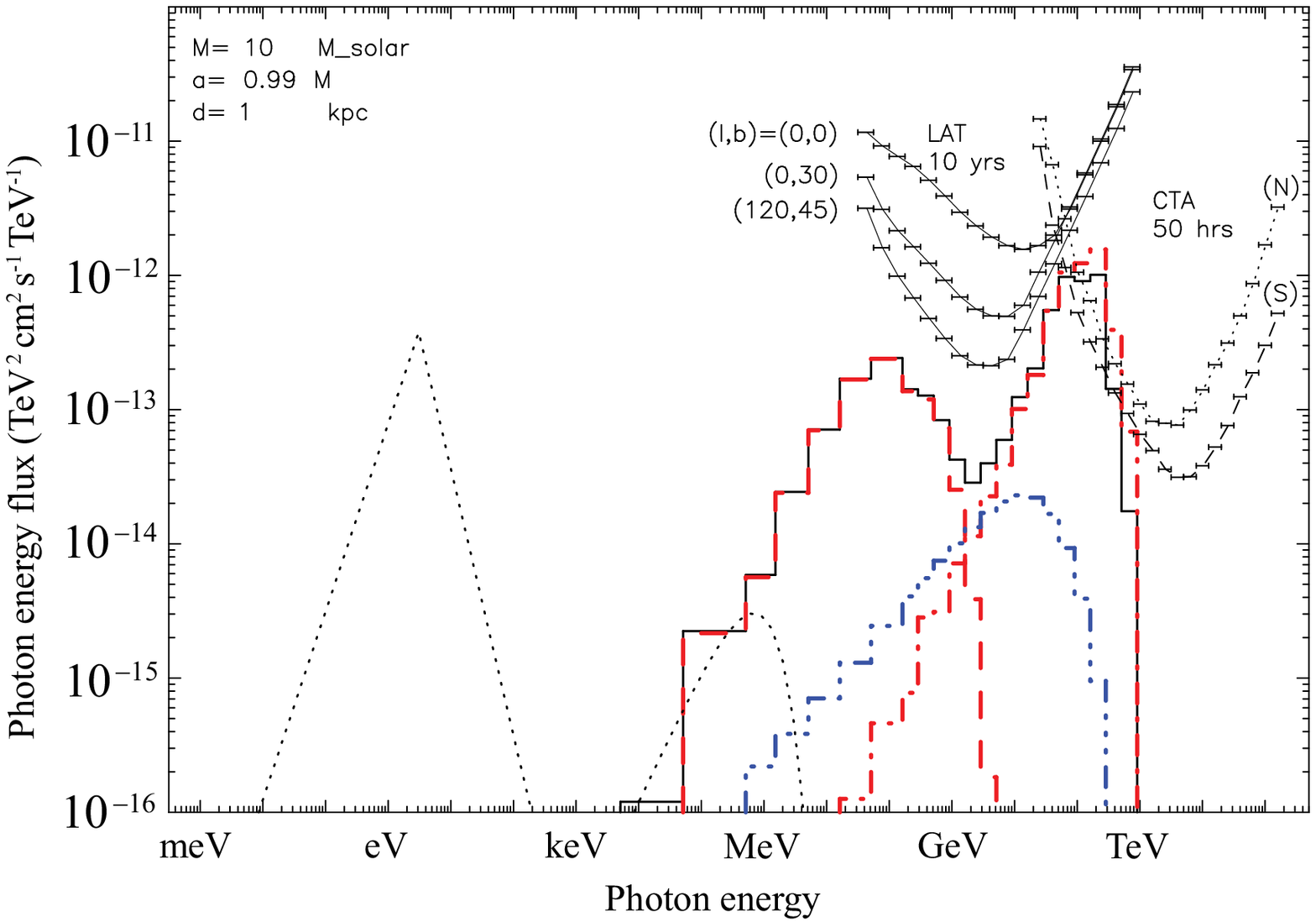}
  \caption{
Similar to the top panel of figure~4, 
but the individual emission components are shown;
$\dot{m}=10^{-4}$ is adopted.
The black solid line represents the same spectrum as the black solid line
in the top panel of figure~4.
The magenta dashed line shows the primary curvature component, while
the magenta dash-dotted line shows the primary inverse-Compton component.
The blue dash-dot-dot-dotted line shows % and the purple dotted lines show 
the sum of the synchrotron and inverse-Compton components
emitted by the secondary pairs.
The thin dotted line shows the input ADAF spectrum.
          }
\label{fig:S4}
\end{figure}

In figure~\ref{fig:S4},
we depict the individual emission components,
selecting the case of $\dot{m}=10^{-4.00}$.
We find that the primary curvature component 
(magenta dashed line) 
dominates between 5~MeV and 0.5~GeV, 
while the primary IC component (magenta dash-dotted line)
does above 5~GeV.
The secondary IC component (blue dash-dot-dot-dotted line)
appears between 0.5~GeV and 5~GeV.
The primary IC component suffers absorption above
0.1~TeV.

\subsection{Dependence on the created current}
\label{sec:results_created}
\begin{figure}
  \centering
     \includegraphics[width=15cm]{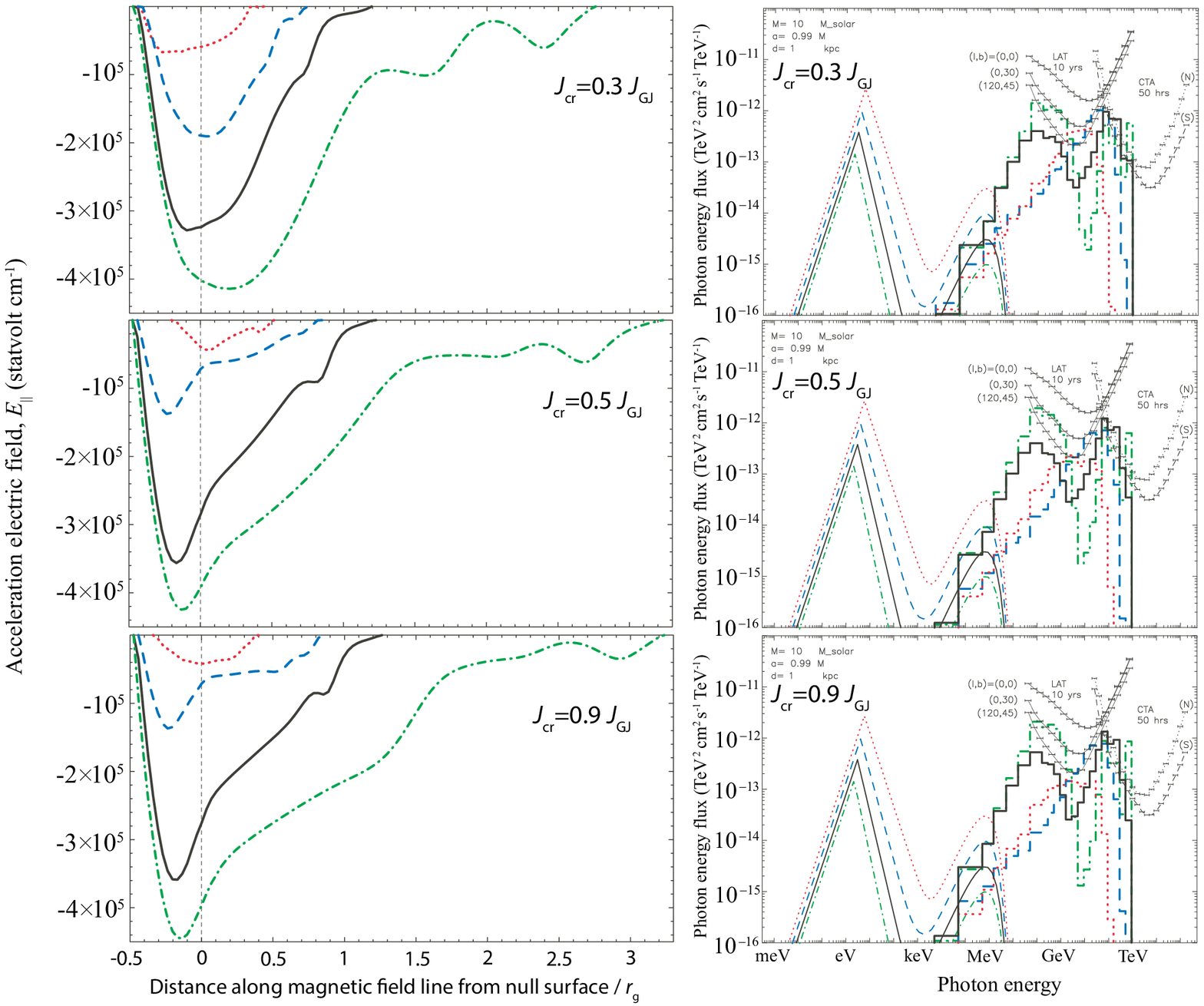}
  \caption{
Dependence of the magnetic-field-aligned electric fields $E_\parallel(s)$ 
(left panels) 
and the SEDs (right panels)
on the created current within the gap.
From the top, the created current density is
30\%, 50\%, and 90\% of the GJ value.
In all the three cases,
the injected current density across the inner or outer boundaries
are set to be zero.
The curves corresponds to the accretion rate of
$\dot{m}=3.16 \times 10^{-4}$ (red dotted), 
$1.77 \times 10^{-4}$ (blue dashed), 
$1.00 \times 10^{-4}$ (black solid), and 
$5.62 \times 10^{-5}$ (green dash-dotted).
          }
\label{fig:Ell_SED_Jcr}
\end{figure}

Let us demonstrate that 
the gap solution exists in a wide range of the created 
electric current within the gap,
$j_{\rm cr} \equiv J_{\rm cr}/J_{\rm GJ}$, 
and that the resultant gamma-ray spectrum little depends on $j_{\rm cr}$.
In figure~\ref{fig:Ell_SED_Jcr}, 
we show the solved $E_\parallel(s)$ (left panels) and SEDs (right panels) 
for three discrete $j_{\rm cr}$'s:
from the top, they corresponds to 
$j_{\rm cr}=0.3$, $0.5$, and $0.9$.
The case of $j_{\rm cr}=0.7$ is presented as
figure~\ref{fig:Epara_mdot} and the top panel of 
figure~\ref{fig:SED_spin}.
It is clear that the gap spectra modestly depends on 
the created current within the gap
as long as the created current is sub-GJ.
Note that there exists no stationary gap solutions 
if the created current density is set to be super-GJ, 
$j_{\rm cr} > 1$.

\subsection{Dependence on the injected current}
\label{sec:results_injected}
\begin{figure}
  \centering
     \includegraphics[width=15cm]{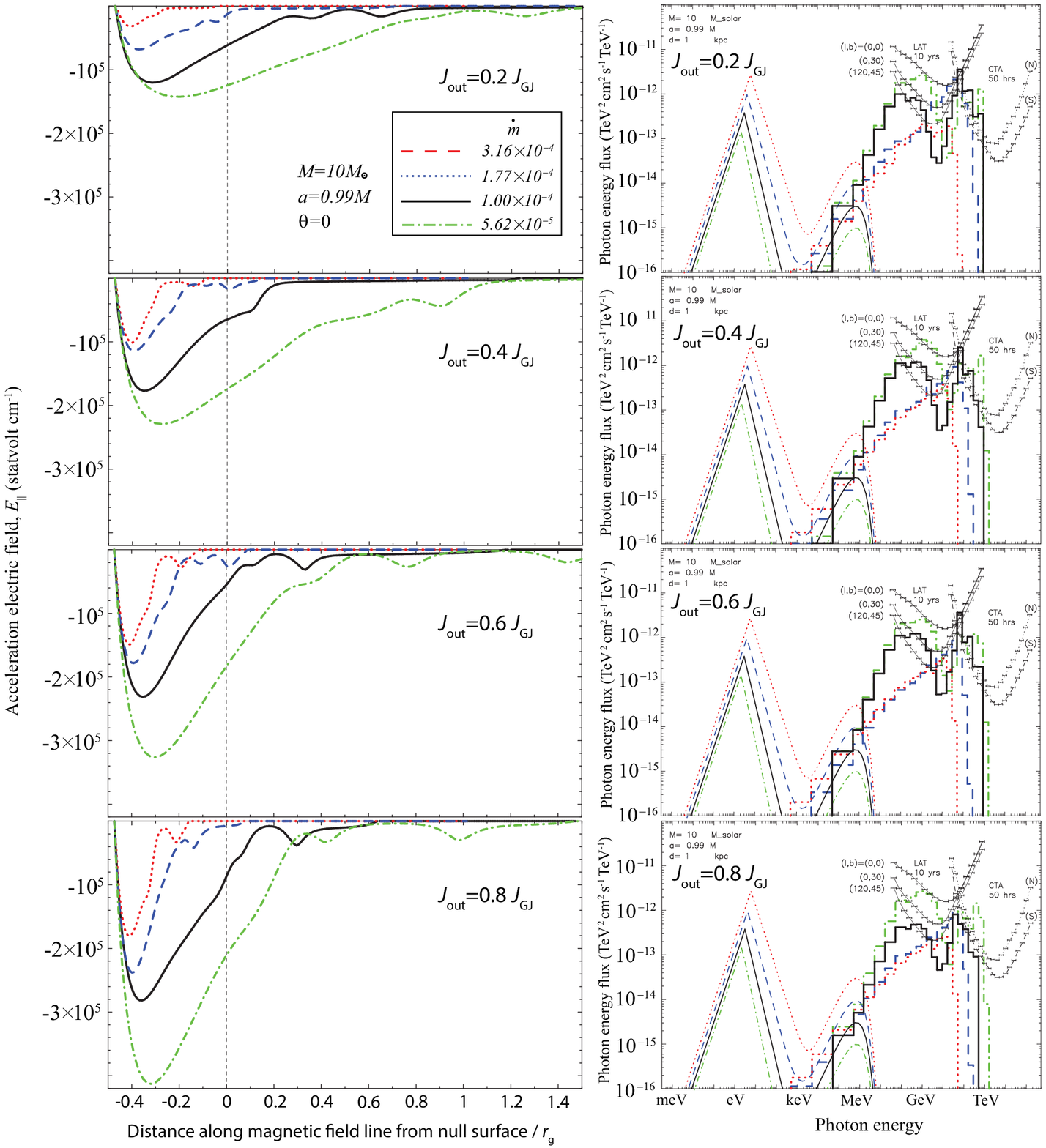}
  \caption{
Dependence of the magnetic-field-aligned electric fields $E_\parallel(s)$ 
(left panels) 
and the SEDs (right panels)
on the injected positronic current across the outer boundary.
From the top, the injected current density is
20\%, 40\%, 60\%, and 80\% of the GJ value.
In all the four cases,
the created current density within the gap 
is fixed to be 70\% of the GJ value,
and the injected electronic current across the inner boundary
are set to be zero.
The curves corresponds to the accretion rate of
$\dot{m}=3.16 \times 10^{-4}$ (red dotted), 
$1.77 \times 10^{-4}$ (blue dashed), 
$1.00 \times 10^{-4}$ (black solid), and 
$5.62 \times 10^{-5}$ (green dash-dotted).
Note that the ordinate scale is ten times greater than 
figure~1 % \ref{fig:Epara_mdot}.
          }
\label{fig:SED_Jout}
\end{figure}

The gap solution exists in a wide range of the injected 
electric currents across the inner and outer boundaries.
In this section, we consider only the current injected across
the outer boundary, 
because the positrons created below the separation surface
(fig.~2 of \citet{hiro16a}, hereafter HP16) may enter the gap
across the outer boundary, 
whereas the electrons created below the inner boundary will fall onto
the horizon.
In the left panels of figure~\ref{fig:SED_Jout}, 
$E_\parallel$ is plotted as a function of 
the Boyer-Lindquist radial coordinate from the null-charge surface. 
The red dotted, blue dashed, black solid, and green dash-dotted
curves corresponds to the cases of 
$\dot{m}= 10^{-3.50}$, $10^{-3.75}$, $10^{-4.00}$, and $10^{-4.25}$, 
respectively.
From the top, each panel show the results for 
$j_{\rm out}=0.2$, $0.4$, $0.6$, and $0.8$. 
In the right panels, SEDs are presented for the same set of 
$j_{\rm out}$ values.
The case of $j_{\rm out}= 0$ is presented as
figure~\ref{fig:Epara_mdot} and the top panel of 
figure~\ref{fig:SED_spin}.

It follows from the left panels that 
$E_\parallel$ shifts inwards with increasing $J_{\rm out}$ 
and that $\vert E_\parallel \vert$ increases
as the gap position shifts inwards,
as suggested by the outer-gap solutions for rotation-powered pulsars
\citep{hiro01b,hiro01c,hiro02}.
Note that the photon-photon collision mean-free path does not decrease
as the gap approaches the horizon,
which forms a striking contract from pulsars.
As a result, the potential drop also increases as the gap shifts inwards
in the case of BHs.
(In the case of pulsars, the soft photons are emitted from the
cooling neutron star surface. 
Thus, as the gap approaches the stellar surface, 
the head-on collisions of inward $\gamma$-rays and 
the outward surface X-rays become more efficient,
decreasing the gap longitudinal size and hence the potential drop.)
In the present BH cases,
the increased $\vert E_\parallel \vert$ and the potential drop
leads to an increased gap luminosity and photon energies
in the local reference frame.
However, due to the gravitational redshift,
the photon energy reduces for a distance static observer.
Accordingly, as the right panels show,
the final SEDs little change if $j_{\rm out}$ changes from $0$
(top panel of fig.~\ref{fig:SED_spin})
to $0.8$ (bottom right panel of this fig.~\ref{fig:SED_Jout}),
although the $E_\parallel(s)$ distribution changes significantly
as figure~\ref{fig:Epara_mdot} and 
left panels of figure~\ref{fig:SED_Jout} show.

% \begin{figure}
%   \epsscale{1.0}
% %  \epsscale{0.50}
%   \plotone{f_1e1_a9900_Ell_SED_4Jout.eps}
%  % \plotone{f2_bw.eps}
%  % \includegraphics[angle=0,scale=0.60]{f5.eps}
%  % \includegraphics[angle=-90,scale=0.60]{fig_5.eps}
%  \caption{
% Dependence of the magnetic-field-aligned electric fields $E_\parallel(s)$ 
% (left panels) 
% and the SEDs (right panels)
% on the injected positronic current across the outer boundary.
% From the top, the injected current density is
% 20\%, 40\%, 60\%, and 80\% of the Goldreich-Julian value.
% The curves corresponds to the accretion rate of
% $\dot{m}=3.16 \times 10^{-4}$ (red dotted), 
% $1.77 \times 10^{-4}$ (blue dashed), 
% $1.00 \times 10^{-4}$ (black solid), and 
% $5.62 \times 10^{-5}$ (green dash-dotted).
% Note that the ordinate scale is ten times greater than 
% figure~1 % \ref{fig:Epara_mdot}.
%  %
%  % \citep[see][]{heiles03}. Plots for all sources are available
%  % in the electronic edition of {\it The Astrophysical Journal}.
%  \label{fig:SED_Jout}
%  }
%  \end{figure}

\section{Discussion}
\label{sec:discussion}
To sum up, we examined stationary solutions of 
the electron-positron accelerator exerted in a rotating BH
magnetosphere.
Depending on the molecular hydrogen density and the BH velocity
with respect to the molecular cloud,
the Bondi accretion rate can be adjusted as
$6\times 10^{-5} < \dot{m} < 10^{-4}$
in the Eddington unit.
In this case, $\gamma$-rays are efficiently emitted outward
in the polar region, typically $\theta \le 15^\circ$,
and the emission from a stellar-mass BH
could be marginally detectable with CTA,
provided that the BH is rapidly rotating (e.g., $a>0.9M$),
and that the accretion rate is adjusted in the range
$6\times 10^{-5} < \dot{m} < 10^{-4}$.
The final photon spectrum little depends on the created current density
within the gap, or on the externally
injected current density across the outer boundary.

\subsection{Comparison with other gamma-ray emission models
from molecular clouds}
\label{sec:disc_comarison}
We compare the gamma-ray emission scenarios from molecular clouds 
(table~\ref{tbl-1}).
{\it In the protostellar jet scenario} \citep{bosch10}, 
jets are ejected from massive protostars
to interact with the surrounding dense molecular clouds,
leading to an acceleration of electrons and protons 
at the termination shocks.
Accordingly, the size of the emission region becomes comparable to 
the jet transverse thickness at the shock. 
{\it In the hadronic cosmic ray scenario} \citep{ginz64,blandford87}, 
protons and helium nuclei are accelerated in the supernova shock fronts,
a portion of which propagate into dense molecular clouds.
As a result of the proton-proton (or nuclear) collisions,
neutral pions are produced and decay into gamma-rays,
whose spectrum becomes a single power-law between 0.001-100 TeV.
Since this interaction takes place most efficiently in a dense gaseous region,
the size of the gamma-ray image will become comparable to the core of 
a dense molecular cloud. 
{\it In the leptonic cosmic ray scenario}
\citep{aharon97,swaluw01,hillas98}, 
electrons are accelerated at 
pulsar wind nebulae or shell-type supernova remnants, 
and radiate radio/X-rays and gamma-rays via synchrotron and
IC processes, respectively.
Since the cosmic microwave background radiation provides 
the main soft photon field in the interstellar medium, 
the size may be comparable to the plerions, 
whose size increases with the pulsar age. 
{\it In the BH-gap scenario}
(see \S~\ref{sec:intro} for references), 
emission size does not exceed $10 r_{\rm g}$. 
Since the angular resolution of the CTA is about 
five times better than the current IACTs, 
we propose that we can discriminate the present BH-gap scenario 
from the three above-mentioned scenarios 
by comparing the gamma-ray image and spectral properties. 
Namely, if a VHE source has a point-like morphology like 
HESS J1800-2400C in a gaseous cloud (\S~\ref{sec:VHE_obs}), 
and if the spectrum has two peaks around 0.01--1 GeV and 0.01--1 TeV, 
but shows (synchrotron) power-law component in neither radio 
nor X-ray wavelengths, we consider that the present scenario 
accounts for its emission mechanism.

\begin{table}
\begin{center}
\caption{Gamma-ray emission models from molecular 
         clouds${}^{\rm a}$.\label{tbl-1}}
% \begin{tabular}{crrrrrrrrrrr}
 \begin{tabular}{clcc}
 \tableline\tableline
 Model & Emission processes (spectral shape; energy range) 
 & Size (cm) \\ 
% & References \\
%% Star & Height & $d_{x}$ & $d_{y}$ & $n$ & $\chi^2$ & $R_{maj}$ & $R_{min}$ &
%% \multicolumn{1}{c}{$P$\tablenotemark{a}} & $P R_{maj}$ & $P R_{min}$ &
%% \multicolumn{1}{c}{$\Theta$\tablenotemark{b}} \\
 \tableline
 Protostellar jets 
 & $e^-$ synchrotron 
    (power-law; $10^{-6}$~eV--$10^2$~eV);
 & $10^{16}$--$10^{17}$ \\
% & 22 \\
   \  
 & $e^-$ Bremsstrahlung 
    (power-law; 0.1~MeV--TeV);
 & \ \\
% & \ \\
   \  
 & $pp$ collisions, $\pi^0$ decays${}^{\rm b}$ % \talenotemark{a}
    (power-law; GeV--TeV)
 & \ \\
% & \ \\
 \tableline
 Cosmic ray hadrons
 & $pp$ collisions, $\pi^0$ decays${}^{\rm b}$ % \talenotemark{a}
    (power-law; GeV--100~TeV)
 & $10^{18}$--$10^{19}$ \\
% & 23,24 \\
 \tableline
 Cosmic ray leptons${}^{\rm c}$
 & $e^-$ synchrotron 
    (power-law; $10^{-6}$~eV--$10^2$~eV);
 & $10^{18}$--$10^{20}$ \\ 
% & 25-27 \\
   \ 
 & $e^-$ IC scatterings
    (broad peak; GeV--10~TeV)
 & \ \\
% &  \\
 \tableline
 BH gap
 & $e^-$ curvature process
    (broad peak; 0.01~GeV--1~GeV);
 & $10^7$ \\ 
% & 14,28-29 \\
   \ 
 & $e^-$ IC scatterings
    (sharp peak; around 0.1~TeV)
 & \ \\
% & \ \\

% 1 &33472.5 &-0.1 &0.4  &53 &27.4 &2.065  &1.940 &3.900 &68.3 &116.2 &-27.639\\
% 2 &27802.4 &-0.3 &-0.2 &60 &3.7  &1.628  &1.510 &2.156 &6.8  &7.5 &-26.764\\
% 3 &29210.6 &0.9  &0.3  &60 &3.4  &1.622  &1.551 &2.159 &6.7  &7.3 &-40.272\\
% 5 & 9607.4 &-0.4 &-0.4 &60 &1.4  &1.669\tablenotemark{c}  &1.574 &2.343 &8.0  
% &8.9 &-33.417\\
% 6 &31638.6 &1.6  &0.1  &39 &315.2 & 3.433 &3.075 &7.488 &92.1 &25.3 &-12.052\\
 \tableline
 \end{tabular}
%% Any table notes must follow the \end{tabular} command.
 \tablenotetext{a}{See \S~6.1 for references.}
 \tablenotetext{b}{Neutral pion ($\pi^0$) decays 
follow proton-proton collisions.}
 \tablenotetext{c}{Gamma-rays emitted by cosmic-ray leptons 
may not be associated with molecular clouds. 
Nevertheless, it is one of the main scenarios of the VHE emissions 
from massive-star forming regions.}
%
% \tablecomments{We can also attach a long-ish paragraph of explanatory
% material to a table.}
 \end{center}
 \end{table}

\subsection{Current injection and time dependence}
\label{sec:disc_injection}
Although the magnetic (i.e., one-photon) pair production is 
also taken into account,
most electron-positron pairs are found to be
produced via photon-photon (i.e., two-photon) collisions,
which take place via two paths.
One path is through the collisions of the two MeV photons 
both of which were emitted from the equatorial ADAF.
Another path is through the collisions of TeV and eV photons;
the former photons were emitted by the gap-accelerated leptons
via inverse-Compton process, 
while the latter were emitted from the ADAF via synchrotron process.
There is, indeed, the third path,
in which the gap-emitted GeV curvature photons collide with
the ADAF-emitted keV inverse-Compton photons;
however, this path is negligible particularly when $\dot{m} \ll 1$.
If the pairs are produced via TeV-eV collisions 
(i.e., via the second path) outside the gap outer boundary, 
they have outward ultra-relativistic momenta
to easily \lq climb up the hill' of the potential $k_0$
(see fig.~2 of HP16) and propagate to large distances
without turning back.
However, if the pairs are produced via MeV-MeV collisions
(i.e., via the first path),
they are produced with sub-relativistic outward momenta;
thus, they eventually return to fall onto the horizon
due to the strong gravitational pull inside the separation surface
(fig.~2 of HP16).
When the returned pairs arrive the gap outer boundary,
only positrons can penetrate into the gap
because of $E_\parallel<0$.
Accordingly, electrons accumulate at the boundary,
whose surface charge leads to the jump of the normal derivative of 
$E_\parallel$.
Thus, although the stationary gap solutions show that the $\gamma$-ray
spectrum little depends on the injected current density
(\S~\ref{sec:results_injected}),
the gap solution inevitably becomes time-dependent due to the
increasing discontinuity of $\vert dE_\parallel/dr \vert$ 
with an accumulated surface charge (in this case, electrons)
at the outer boundary.
If the injected current is much small compared to the GJ current, 
the time dependence will be mild.
However, if the injected current becomes a good fraction of the 
GJ current, the assumption of the stationarity becomes invalid,
as pointed out by \citet{levi17}.
In this sense, 
a caution should be made in the applicability of the stationary
solutions presented in this paper, when the injected current
is non-negligible compared to the current created within the gap.

\subsection{Stability of stationary black hole gaps}
\label{sec:disc_stability}
In this subsection,
we consider the stability of our stationary gap solutions.
In the case of pulsar polar caps, it has been revealed
that the pair production cascade takes place in a highly time-dependent
way by particle-in-cell (PIC) simulations 
\citep{timokhin13,timokhin15,timokhin10}.
Moreover, in the case of BH magnetospheres, it is recently 
demonstrated that a gap exhibits rapid spatial and temporal oscillations 
of the magnetic-field-aligned electric field and current 
by 1-D PIC simulations \citep{levi17,cheng18}.
It is, however, out of the scope of the present paper to perform
a PIC simulation or a linear perturbation analysis to examine
the stability of a stationary gap solution.
Instead, we will qualitatively discuss why we seek stationary 
solutions, comparing with pulsar outer-magnetospheric and polar-cap gaps.

We start with discussing the pulsar outer (-magnetospheric) gaps,
because they have essentially the same electrodynamics as BH gaps. 
Since the neutron star's dipole magnetic field lines have convex
geometry, the magnetic field becomes perpendicular with respect to
the star's rotation axis at a good fraction of the so-called
\lq\lq light cylinder radius.'' 
In this case, if the magnetosphere is highly vacuum in the sense
$\vert \rho \vert \ll \vert \rho_{\rm GJ} \vert$
in equation~(\ref{eq:pois}),
$\rho-\rho_{\rm GJ} \approx -\rho_{\rm GJ} > 0$ (or $<0$) holds
in the lower (or the upper) half of the gap.
As a result, $E_\parallel$ has a positive (or negative)
gradient in the lower (or the upper) half;
thus, the acceleration electric field naturally closes.
Note that $E_\parallel>0$ is realized when the magnetic axis 
resides in the same hemisphere as the rotation axis.
Without loss of any generality, we can adopt such a positive $E_\parallel$
in the outer-gap model.

\begin{figure}
  \centering
     \includegraphics[width= 7.3 truecm]{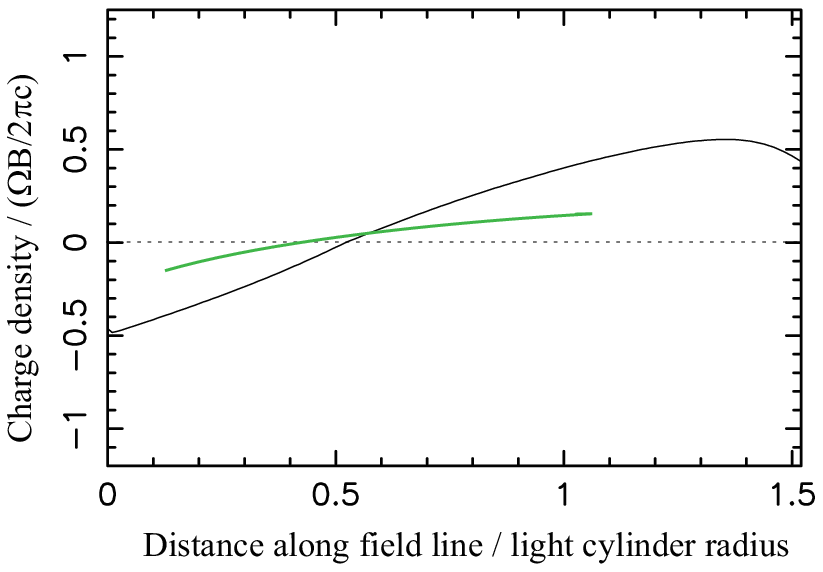}
     \includegraphics[width= 7.3 truecm]{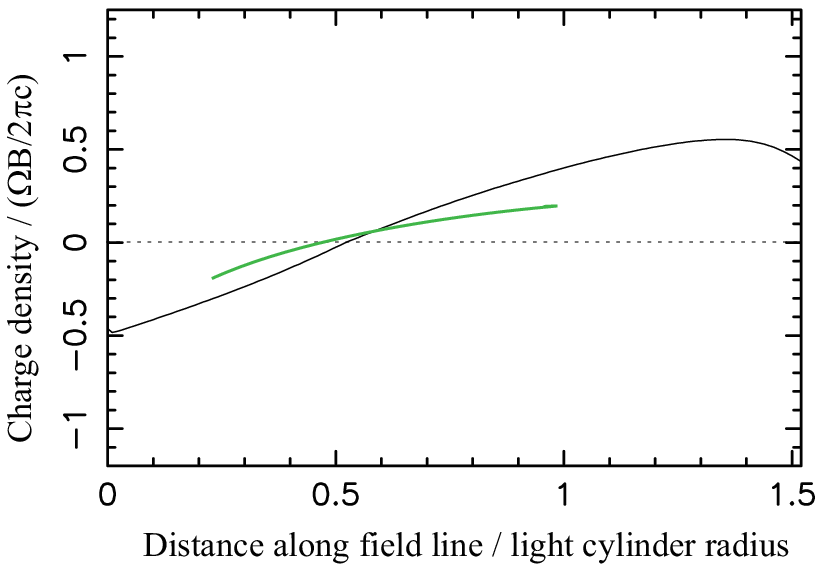}
  \caption{
Schematic picture of the distribution of 
a dimensionless charge density per magnetic flux tube, 
$(2\pi/\Omega_{\rm F})(\rho/B)$,
in the pulsar {\it outer-gap} model.
A {\it positive} acceleration electric field, $E_\parallel$, arises,
because $\rho-\rho_{\rm GJ}>0$ holds (i.e., $dE_\parallel/ds>0$),
in the inner part of the gap, and 
because $\rho-\rho_{\rm GJ}<0$ holds (i.e., $dE_\parallel/ds<0$) 
in the outer part.
The dimensionless Goldreich-Julian charge density, 
$(2\pi/\Omega_{\rm F})(\rho_{\rm GJ}/B)$ is depicted by the solid curve.
{\it Left:}
As an initial state, we consider a typical $\rho/B$ distribution
(green solid curve) of a non-vacuum outer gap.
Created positrons (or electrons) are accelerated outwards (or inwards)
by the positive $E_\parallel$;
therefore, $\rho/B$ has a positive gradient.
{\it Right:}
Imagine that the pair production increases perturbatively.
Because $E_\parallel>0$, created and migrated positrons
(or electrons) increase (or decrease) $\rho$ at the
outer (or inner) part of the gap.
As a result, the reduced $\rho-\rho_{\rm GJ}$ partly cancels
the original $E_\parallel$, reducing the perturbatively increased
pair production.
Because of this {\it negative} feedback effect,
the outer-gap solution depends on the pulsar parameters,
as well as the magnetospheric current, only modestly \citep{hiro01}.
       }
\label{fig_rho_OG_1}
\end{figure}

Because $E_\parallel>0$, positrons (or electrons) are accelerated
outwards (or inwards).
Thus, $\rho$ has a positive gradient along individual 
magnetic field lines.
When the gap closure condition is satisfied, there exists a stationary
solution whose $\rho/B$ distribution can be illustrated 
as the left panel of figure~\ref{fig_rho_OG_1}.
If pair production increases perturbatively from this stationary solution,
$\rho-\rho_{\rm GJ}$ decreases its absolute value, 
which leads to an decrease of $E_\parallel$
due to the reduced gradient of $\vert \rho-\rho_{\rm GJ} \vert$, 
and hence an decrease of the pair production 
(right panel of fig.~\ref{fig_rho_OG_1}).
Note that this negative feedback effect works because of
$E_\parallel>0$, which stems from the fact that there exists 
a null-charge surface in the gap.
As a result, the outer gap solutions exist for a wide range of 
pulsar parameters such as the period, period derivative, 
neutron-star surface temperature, 
inclination of the magnetic axis with respect to the star's rotation axis,
as well as the magnetospheric current,
from young, middle-aged to millisecond pulsars.
Analogous (but still qualitative) argument of gap stability is possible 
if we use the gap closure condition;
see section 5.2 of \citet{hiro01} for details.

Because null-charge surfaces also exist around rotating BHs,
the gap electrodynamics little changes between pulsar outer gaps and
BH gaps.
Around a rotating BH, a null surface is formed 
near the event horizon by the frame-dragging.
However, around a rotating neutron star, it is formed 
in the outer magnetosphere (i.e., far away from the neutron star) 
by the convex magnetic-field geometry.
Accordingly, a {\it negative} feedback effect also works in BH gaps.
That is, the gap closure condition,
which is required for a BH gap to be stationary,
is accommodated for a wide range of magnetospheric current values,
as explicitly demonstrated in \S~\ref{sec:BH_gap_results}.

\begin{figure}
  \centering
     \includegraphics[width= 7.3 truecm]{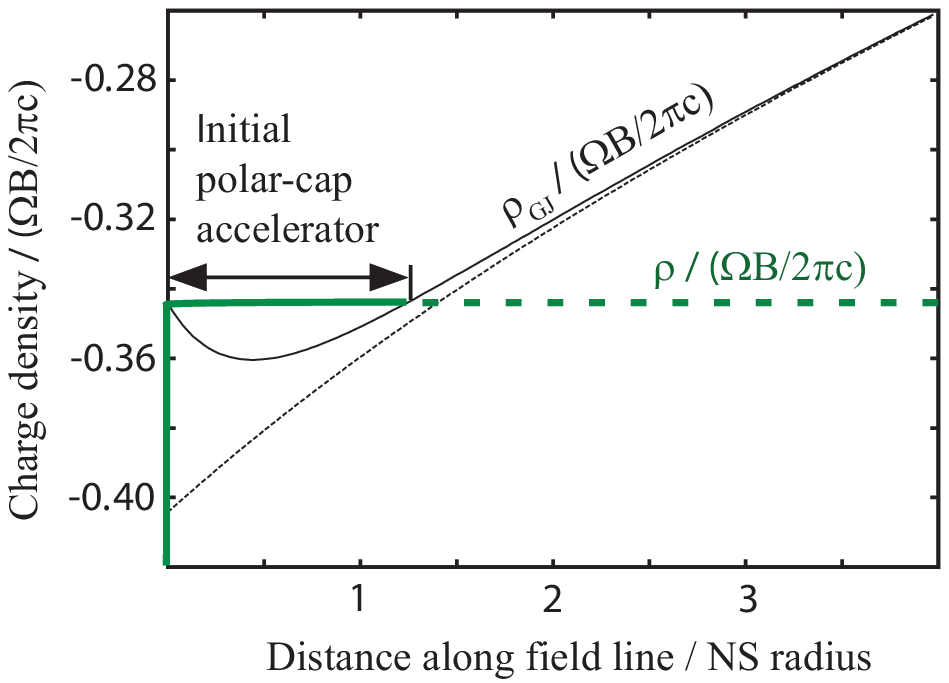}
     \includegraphics[width= 7.3 truecm]{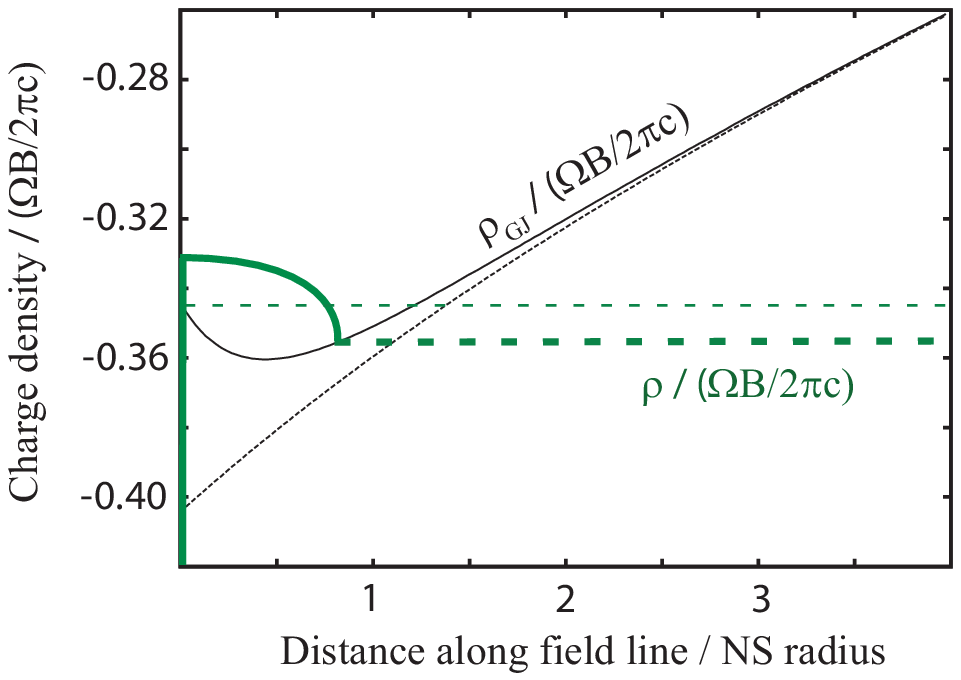}
  \caption{
Schematic picture of the distribution of a
$\rho/B$ in the pulsar {\it polar-cap} model.
A {\it negative} $E_\parallel$ arises,
because $\rho-\rho_{\rm GJ}>0$ holds in the entire gap except for 
the direct vicinity of the neutron star surface.
This negative $E_\parallel$ extract electrons from the stellar surface
as a space-charge-limited flow.
The dimensionless Goldreich-Julian charge density, 
$\rho_{\rm GJ}/B$ is depicted by the solid curve, while its Newtonian
value is plotted by the thin black dotted curve.
{\it Left:}
As an initial state, we consider no pair production;
thus, $\rho/B$ tends to a constant value, 
as the horizontal green line shows.
The polar-cap accelerator (i.e., gap) arises approximately 
within the region where $\rho > \rho_{\rm GJ}$ holds 
above the stellar surface.
{\it Right:}
The outwardly accelerated electrons emit the $\gamma$-rays
that materialize as pairs in the upper-most part of the gap.
The separated electrons move outwards reducing $\rho$ from its
initial value, while positrons move inwards increasing $\rho$.
Accordingly, in the lower part of the initial gap, 
increased $E_\parallel$ leads to a further enhanced pair production;
this {\it positive} feedback effect results in an instability
\citep{timokhin13}.
          }
\label{fig_rho_PC_1}
\end{figure}

Since there frequently appears a confusion between pulsar polar-cap
and outer-gap (and hence BH-gap) electrodynamics,
particularly on the stability argument,
it is helpful to describe also the pulsar polar-cap accelerator.
In a pulsar polar cap, there exists no null surface.
Thus, if a three-dimensional polar cap region is charge starved in the sense
$\vert \rho-\rho_{\rm GJ}\vert \ll \rho_{\rm GJ}$,
a positive $-\rho_{\rm GJ}$ leads to a {\it negative} $E_\parallel$
when the magnetic axis resides in the same hemisphere as the rotation axis.
Accordingly, electrons are drawn from the neutron star surface
as a space-charge-limited flow.
In the direct vicinity of the neutron star surface,
the non-relativistic electrons produces a large negative $\rho$
(the green vertical lines along the ordinates in figure~\ref{fig_rho_PC_1})
such that $\rho-\rho_{\rm GJ} \approx \rho < 0$.
In the outer part of a polar cap accelerator, on the other hand,
relativistic electrons produces a moderate negative charge density
such that $\rho-\rho_{\rm GJ} > 0$ 
(left panel of fig.~\ref{fig_rho_PC_1}).
Thus, $E_\parallel$ has a negative gradient
along the magnetic field in the direct vicinity of the neutron star,
but it has a positive gradient
near the upper boundary where $\rho-\rho_{\rm GJ} \approx 0$.
Accordingly, we obtain a negative $E_\parallel$ in the pulsar polar caps.

However, if a small-amplitude pair production takes place in the polar gap,
inwardly migrating positrons (or outwardly migrating electrons)
result in an increased (or decreased) $\rho-\rho_{\rm GJ}$ 
in the lower (or the upper) half of the gap 
(right panel of fig.~\ref{fig_rho_PC_1}).
Accordingly, $\vert E_\parallel \vert$ increases (or decreases)
in the lower part (or the upper-most part).
The increased $\vert E_\parallel \vert$ further enhances pair production
in the upper-most part,
and $\vert E_\parallel \vert$, and hence the pair production increases 
with time.
Because of this {\it positive} feedback effect, instability sets in,
as demonstrated by PIC simulations 
\citep{timokhin13,timokhin10}.
In short, pulsar polar cap accelerators are inherently unstable for 
pair production 
because of $E_\parallel < 0$,
which stems from the fact that there exists no null-charge surface
in the pulsar polar-cap region.

On these grounds, we cannot readily conclude that the stationary 
BH gap solutions presented in this paper are unstable 
because of the highly time-dependent nature of pulsar polar gaps.
A careful examination with a PIC simulation is needed for BH gaps,
as performed recently by \citet{levi17}.
Since the saturated solution given by \citet{levi17} 
is much less violently
time-dependent compared to pulsar polar-cap accelerators \citep{timokhin13},
we have sought stationary solutions of BH gaps as the first step
in the present paper.

%% If you wish to include an acknowledgments section in your paper,
%% separate it off from the body of the text using the \acknowledgments
%% command.

%% Included in this acknowledgments section are examples of the
%% AASTeX hypertext markup commands. Use \url without the optional [HREF]
%% argument when you want to print the url directly in the text. Otherwise,
%% use either \url or \anchor, with the HREF as the first argument and the
%% text to be printed in the second.

\acknowledgments
This work is supported by the Theoretical Institute for 
Advanced Research in Astrophysics (TIARA) operating under Academia Sinica, 
and by their High Performance Computing system. 
SM is supported by the Ministry of Science and Technology (MoST) 
of Taiwan, MoST 103-2112-M-001-032-MY3 and 106-2112-M-001-011, and 
AKHK through 105-2112-M-007-033-MY2 and 106-2628-M-007-005.

\end{document}